\documentclass[a4paper,10pt]{article}
\usepackage{graphicx}
\usepackage{amssymb}
\usepackage{amsmath}
\usepackage{nicefrac}
\usepackage{dcolumn}
\usepackage{multirow}
\usepackage{cite}
\usepackage{mathrsfs}
\usepackage{bm}
\usepackage[usenames,dvipsnames]{xcolor}
\usepackage[colorlinks=true,citecolor=Blue,linkcolor=RubineRed,urlcolor=Blue]{hyperref}
\usepackage{bbm}
\usepackage{hyperref}
\usepackage{comment}
\usepackage{amsfonts}
\usepackage{float}
\usepackage{caption}
\usepackage{subcaption}

\usepackage{xcolor}
\usepackage{soul}
\setstcolor{red}

\begin{document}

\huge

\begin{center}
Plasma density effects on the electron impact ionization
\end{center}

\vspace{0.5cm}

\large

\begin{center}
Djamel Benredjem$^{a,}$\footnote{djamel.benredjem@universite-paris-saclay.fr}, Jean-Christophe Pain$^{b,c}$, Annette Calisti$^{d}$ and Sandrine Ferri$^{d}$
\end{center}

\normalsize

\begin{center}
\it $^a$Universit\'e Paris-Saclay, CNRS, Laboratoire Aim\'e Cotton, F-91405 Orsay, France\\
\it $^b$CEA, DAM, DIF, F-91297 Arpajon, France\\
\it $^c$Universit\'e Paris-Saclay, CEA, Laboratoire Mati\`ere sous Conditions Extr\^emes,\\
\it 91680 Bruy\`eres-le-Ch\^atel, France\\
\it $^d$Aix-Marseille Université, CNRS, Physique des Interactions Ioniques et Moléculaires, UMR7345, Campus Saint Jerome, 13397 Marseille Cedex 20, France
\end{center}

\vspace{0.5cm}

\begin{abstract}
We present new results on the ionization by electron impacts in a dense plasma. We are interested in the density effect known as the ionization potential depression and in its role in atomic structure. Rather than using the well-known Stewart-Pyatt or Ecker-Kr\"oll formulas for the ionization potential depression, we consider a distribution function of the ionization energy, which involves the plasma fluctuations due to ion dynamics. This distribution is calculated within classical molecular dynamics. The removal of the noise yields a new distribution which is composed of a small set of Gaussian peaks among which one peak is selected by considering the signal-to-noise ratio. This approach provides an ionization potential depression in good agreement with experimental results obtained at the Linac Coherent Light Source facility. Our results are also compared to other calculations.

In a second part, we investigate the effects of the ionization potential depression and the fluctuations on ionization by electron impacts. We propose a new expression of the cross section, based on an average over the ionization energy distribution. This cross section can be calculated analytically. The main strength of our work is to account for the fluctuations due to ion dynamics.
\end{abstract}

\section{Introduction}

The knowledge of the radiative properties of dense and hot plasmas requires accurate cross sections for the processes involving an ion and free electrons. Different methods were used to calculate the cross sections of various elements. For instance, Colgan \textit{et al.} studied the excitation and ionization of Si, Cl and Ar, in the context of magnetic fusion and astrophysical modeling \cite{Colgan2008}. The ionization of Ne and Au in hot and dense plasmas was investigated by Pindzola \textit{et al.} \cite{Pindzola2008}. At high density, the atomic processes are affected by the plasma environment of the radiator, which may cause level shifts \cite{Belkhiri2015} or continuum lowering. In both cases, the cross section can be significantly modified \cite{Fontes1993}.

In this work, we are interested in the ionization of aluminum by electron impacts. In a recent work, we calculated the ionization cross section of aluminum in a dense plasma \cite{Benredjem2022}. We accounted for the ionization potential depression (IPD) because its effect is important in high-density plasmas. The IPD was investigated decades ago, and useful formulas were established (see Refs. \cite{Stewart-Pyatt1966,Ecker-Kroll1963}). A new interest in this effect is illustrated by theoretical works on aluminum \cite{Crowley2014,Iglesias2013}, silicon \cite{Zeng2020} and iron \cite{Zeng2022}, and measurements. Two experiments were carried out at the Linac Coherent Light Source facility (LCLS, Stanford). The first one \cite{Ciricosta2016} measured the IPD of aluminum and magnesium for several ion charges, and the second one \cite{Kraus2019} focuses on carbon. Another experiment on aluminum was performed at the Orion laser facility \cite{Hoarty2013} at density and temperature in the ranges 1$-$10 g/cm$^3$ and 500$-$700 eV, respectively. While several calculations agree with the experiment of Ciricosta \textit{et al}. \cite{Ciricosta2016} for low ion charges (O-, N- and C-like aluminum), they fail to reproduce the measurements for the highest ion charges (B- and Be-like). This is particularly the case of the Ecker-Kr\"{o}ll formula \cite{Ecker-Kroll1963}. The formula of Stewart and Pyatt \cite{Stewart-Pyatt1966} does not agree with the LCLS measurements but seems to be consistent with the Orion laser experiment.

Although the plasma fluctuations induced by ion charge dynamics are known to be important, their effect on the ionization potential is not well understood. Iglesias and Sterne \cite{Iglesias2013} took into account the fluctuations of the density of free electrons and, consequently, of the ion sphere radius, and proposed simple analytical IPDs within the Stewart-Pyatt \cite{Stewart-Pyatt1966} and Ecker-Kr\"{o}ll \cite{Ecker-Kroll1963} models. Our approach is twofold. We first investigate the effect of the fluctuations on IPD values and then calculate the ionization cross section. The IPD is calculated within the framework of the classical molecular dynamics (CMD) \cite{Calisti2009,Calisti2017}. Owing to the fluctuations, we obtain a distribution of the ionization energy rather than a unique value. 

In order to extract a useful ionization energy from the CMD distribution, we have to remove the numerical noise. The IPD is defined as the difference between the ionization energy of the isolated atom and the one that is deduced from the CMD distribution. Our calculations show a good agreement with experiment, for O- to Be-like aluminum.

Our previous calculations of the electron impact ionization (EII) cross sections \cite{Benredjem2022} rely on the Lotz formula \cite{Lotz1967}, where the ionization energy was assumed to be the average of the ionization energy over the CMD distribution. Unfortunately, in many cases considering isolated ions, the Lotz formula shows substantial differences with accurate calculations obtained with robust codes such as FAC \cite{Gu2008} and HULLAC \cite{BarShalom2001}. In this work, we consider a more suitable formula involving adjustable parameters to allow a very good agreement with both codes, for isolated ions. We also propose to define the cross section as an average over the ionization energy distribution, for each incident electron energy. This is a more satisfactory approach to account for plasma fluctuations in cross-section calculations.

In Sec. \ref{IPD}, we compare the distribution of the ionization energy obtained by the CMD modeling to Gauss, Gram-Charlier \cite{Gram1883,Charlier1905a,Charlier1905b,Kendall1994} and Weibull \cite{Weibull1951,Frechet1927} distributions. In Sec. \ref{Noise Reduction}, we show that removing the noise from the CMD distribution provides a small set of Gaussian peaks. We are then able to select a unique peak which is assumed to be the relevant ionization energy. Finally, the obtained IPD is compared to experimental results \cite{Ciricosta2016} for O- to Be-like aluminum. Section \ref{EII cross section} is devoted to EII cross-section calculations. We use a formula similar to the Kim cross section \cite{Kim1992}, in which the ionization energy is associated to the selected Gaussian peak. The cross section $-$averaged over the Gauss distribution$-$ is calculated analytically. However, the resulting formula involves special functions and is then rather cumbersome. The numerical calculation does not present any difficulty and is therefore preferable. Nevertheless, we propose a simple and accurate analytical expression, based on a cubic-spline representation of the Gaussian.

\section{Ionization potential depression}\label{IPD}

\subsection{Two analytical formulas}

The two well-known IPD formulas are briefly presented. In the calculation of Stewart and Pyatt \cite{Stewart-Pyatt1966}, the IPD of an ion of net charge $ze$ is expressed in terms of the ratio of the Debye length $\lambda_D$ and the ion sphere radius $R_0=[3/(4\pi N_i)]^{1/3}$:
\begin{equation*}
    I_{\rm SP}(z)=\frac{3(z+1)e^2}{2R_0}\left\lbrace\left [ 1+\left (\frac{\lambda_D}{R_0}\right )^3\right ]^{2/3}-\left (\frac{\lambda_D}{R_0} \right )^2\right\rbrace,
\end{equation*}
where $e$ is the electron charge, $N_i$ the ion (number) density. We have set $4\pi\epsilon_0=1$, $\epsilon_0$ being the permittivity of the vacuum. In the literature, one generally uses the high-density limit of the above formula:
\begin{equation*}
    I_{\rm SP-HD}(z)=\frac{3(z+1)e^2}{2R_0}.
\end{equation*}

The Ecker-Kr\"oll formula \cite{Ecker-Kroll1963} reads:
\begin{equation*}
    I_{\rm EK}(z)=\frac{(z+1)e^2}{R_0}\left\{\begin{array}{ll}
    R_0/\lambda_D & \;\;\;\;\mathrm{if} \;\;\;\; N_{\rm cr}\ge N_i(1+\overline{Z})\\
    C(1+\overline{Z})^{1/3} & \;\;\;\;\mathrm{if} \;\;\;\; N_{\rm cr}< N_i(1+\overline{Z}),
    \end{array}\right.
\end{equation*}
where $\bar{Z}$ is the average ion charge and $N_{\rm cr}$ the critical density, given by
\begin{equation*} 
    N_{\rm cr}=\frac{3}{4\pi}\left(\frac{k_{\rm B} T}{Z^2e^2}\right)^3,
\end{equation*} 
where $k_{\rm B}$ is the Boltzmann constant, $T$ the temperature and $Z$ the atomic number. The constant $C$ is determined by imposing the continuity of the IPD at the critical density, yielding
\begin{equation*}
    C=\left[\frac{R_0}{(1+\overline{Z})^{1/3}\lambda_D}\right]_{N_{\rm cr}}.
\end{equation*}
In the present work, we take $C=1$, as in Refs. \cite{Ciricosta2012,Preston2013,Ciricosta2016}.

\subsection{Ionization energy distributions}

\subsubsection{Classical Molecular Dynamics modeling}

The CMD modeling involves an electrically neutral two-component plasma, \textit{i.e.}, a plasma composed of ions of various charges and electrons. Such a task requires a soft ion-electron potential which removes the Coulomb divergence at short distances and accounts for some quantum effects. Within the limits of classical mechanics, all charge-charge interactions are accounted for in the particle motion. The system inside the simulation box is neutral. Periodic boundary conditions are used and the Newton's equations are solved using a Velocity-Verlet algorithm \cite{Verlet1967}. Electron-electron (e-e) or ion-ion (i-i) interactions are taken to be Coulombic:
\begin{equation}
V_{\mathrm{ii,ee}}=Z_{\mathrm{i,e}}^2e^2e^{-r/\lambda}/r.
\end{equation}
The interactions are screened at a distance $\lambda\approx L/2$, where $L$ is the size of the simulation box (typically of the order of a few $\lambda_D$). The number of ions included in the simulation is of the order of 100. Since for aluminum the number of electrons per atom is 13, the total number of particles in the box is of the order of 1400. Molecular dynamics involving opposite charges requires a regularized potential at short and large distances:
\begin{equation}
V_{\mathrm{ie}}(r)=-Z_ie^2e^{-r/\lambda}(1-e^{-r/\delta})/r
\end{equation}
where the regularization distance $\delta$:
\begin{equation}
\delta(Z)=-Z_ie^2/E_i
\end{equation}
is associated to the ionization energy $E_i$ of each ion stage. An electron located at an ion ($r=0$) occupies the fundamental state of the ion whose charge is $Z$ with a nucleus charge $Z+1$. We follow the particle motion with very small time steps $\approx$ 10$^{-20}$ s appropriate for the description of the micro motion of electron around ions. The time steps are small compared to typical collision rates, and thus statistics on collisional events are expensive. The setting up of the population of electrons temporary trapped in the ion wells, \textit{i.e.}, the reaching of the required equilibrium state, depends on collisional events between electrons, and is therefore a very slow process. The choice of the ion-electron regularized potential associated to the knowledge of the position and velocity of individual particles at each time step, allows us to design a collisional ionization/recombination process. Ionization/recombination mechanisms rely on an approximate analysis of collisional events between one ion and one or two electrons. The concept of collisions is not straightforward as the interaction involves all particles within the screening length. The definition of a collisional process is crucial but necessarily empirical.\\
\indent The location of the two nearest-neighbor electrons and the sign of their total energy is used to evaluate if locally the plasma, at one step of its evolution, is favorable to an ionization (positive energy) or a recombination (negative energy) of the ion. This ionization/recombination process implemented in the code has two fundamental functions: it allows for the evolution of the charge state population towards a stationary state depending on temperature, density and composition of the plasma, and favors the setting up of a population of electrons temporary trapped in the ion wells. A preparation phase of the particle set (into the simulation box) before extracting any sampling from simulations, is necessary. At the end of the preparation phase the system follows a quasi-stable evolution with stationary trapped and free-electron populations. \\
\indent The main idea of the model is to extract from the simulated particle positions and velocities, a local characterization of the plasma around an ion I in order to determine if the conditions are favorable to an ionization or recombination of this ion. To this end, the mutual nearest neighbor, ${\rm NN_I}$ and the next nearest neighbor, ${\rm NNN_I}$, electrons of I are identified and tracked. Their total energy is calculated accounting for the complexity of the potential energy surface around I including the ionization energy lowering due to the surrounding charges. A shell noted $S_I$, formed with ${\rm NN_I}$ and ${\rm NNN_I}$, is defined as the nearest environment of I if ${\rm NN_I}$ is localized at a distance $d_I$ of I such as $\delta(Z_I) < d_I < \sqrt{2}\,\delta(Z_I)$. Depending on the total energy of the two neighboring electrons, the shell is labeled ``hot'' (positive energy favorable to ionization) or ``cold'' (negative energy favorable to recombination). A hot or cold shell around an ion undergoes respectively either a pre-ionization (\textit{i.e.}, an increase by 1 of the ion charge and the appearance of one electron localized at the ion), or a recombination (a decrease by 1 of the ion charge and the removal of the nearest-neighbor electron with a transfer of the kinetic energy difference to ${\rm NNN_I}$). The pre-ionized state, \textit{i.e.}, an ion with a trapped electron can then be converted into an ionized state through multiple collisions. In this framework, the ionization will be considered as complete when a new hot shell surrounds the ion opening the way to a further pre-ionization. In the meantime, the ion is considered as excited if the ion potential traps more than one electron. \\ 
\indent It is worth emphasizing that the model does not account for the coupling with radiation and that the density of ionic excited states is replaced by its continuous equivalent (classical approximation). During the initial step of equilibration, the system is brought to equilibrium using a thermostat. Once the system has reached an equilibrium state, the ionization/recombination process becomes scarce as compared to the equilibration step. \\
\indent The time for an electron to cross the average distance $r_0=[3/(4\pi N_e]^{1/3}$, $N_e$ being the electron density, is of the order of $10^{-16}$ s. The relaxation time of the electron velocity distribution is three orders of magnitude larger, \textit{i.e.}, 10$^{-13}$ s. In the present work, the typical run time is about a few hours. Taking advantage of the characteristics of the ionization protocol, it is possible, when the ion is in a pre-ionization state, to measure the energy required to ionize an electron from the ground state of an ion, while taking into account all the interactions with the surrounding plasma. Due to the fluctuating local environment of the ions, the ionization energy is then depicted by a distribution function. The IPD is then written as the difference between the ionization energies of the isolated and the surrounded ion. The latter is inferred from the CMD distribution.\\
\indent A detailed description of the CMD model as well as the capabilities of the code Bingo giving the ionization energy distribution are described in Refs. \cite{Calisti2009,Calisti2017}. In order to characterize and represent the CMD, we have considered the Gauss, Gram-Charlier and Weibull distributions.

\subsubsection{Gauss and Gram-Charlier distributions}

Let us first define the $p-$order moment of the CMD distribution $\mathscr{D}(E_i)$ by
\begin{equation*}
    \mu_{p}=\int E_i^p\,\mathscr{D}(E_i)\,dE_i,
\end{equation*}
and the centered $p$-order moment
\begin{equation*}
    \mu_{p,c}=\int (E_i-\mu_1)^p\,\mathscr{D}(E_i)\,dE_i.
\end{equation*}
The Gauss distribution reads
\begin{equation*}
    {\rm G}(u)=\frac{1}{\sqrt{2\pi\,v}}\exp\left(-u^2/2\right),
\end{equation*}
where $u=(E_i-\mu_1)/\sqrt{v}$ is an adimensional parameter. The quantities $\mu_1$ and $v=\mu_2-\mu_1^2=\mu_{2,c}$ represent the average energy and the variance, respectively. 

The Gram-Charlier distribution is given by
\begin{equation*}
     {\rm GC}(u)=\frac{1}{\sqrt{2\pi\,v}}\exp\left(-u^2/2\right)\left[ \sum_{k=0}^{\infty}c_k~{\rm He}_k\left(\frac{u}{\sqrt{2}}\right)2^{-k/2}\right],
\end{equation*}
where the polynomials ${\rm He}_k$ can be expressed in terms of the Hermite polynomials ${\rm H}_k$, as
\begin{equation*}
    {\rm He}_k(x)=\frac{1}{2^{k/2}}{\rm H}_k\left(\frac{x}{\sqrt{2}}\right).
\end{equation*}
The coefficients $c_k$ are given by 
\begin{equation*}
    c_k=\sum_{j=0}^{\lfloor k/2\rfloor}\frac{(-1)^j}{j!\,(k-2j)!\,2^j}\,\alpha_{k-2j},
\end{equation*}
where $\lfloor\cdot\rfloor$ denotes the integer part. The coefficient $\alpha_k$ is the dimensionless centered $k$-order moment of the distribution:
\begin{equation*}
    \alpha_k=\frac{\mu_{k,c}}{v^{k/2}}=\frac{1}{v^{k/2}}\sum_{p=0}^k\binom{k}{p}\mu_p(-\mu_1)^{k-p}.
\end{equation*}
If we limit ourselves to the fourth order, the distribution can be written as:
\begin{equation*}
    {\rm GC}_4(u)=\frac{1}{\sqrt{2\pi\,v}}\exp\left(-u^2/2\right)\left[ 1-\frac{\alpha_3}{2}\left(u-\frac{u^3}{3}\right)+\frac{\alpha_4-3}{24}\left(3-6u^2+u^4\right)\right],
\end{equation*}
where $\alpha_3$ and $\alpha_4$ are the skewness and the kurtosis, respectively. $\alpha_3$ characterizes the asymmetry of the distribution and $\alpha_4$ its sharpness.

\subsubsection{Weibull distribution}

The characterization and computation of rare events occurring in molecular dynamics are longstanding issues \cite{Hartmann2014}. In the CMD simulations performed with the Bingo code, the ionization/recombination processes are rare events. Therefore, the Gauss distribution (as well as its generalization to include higher-order moments such as the Gram-Charlier expansion series), may not be able to represent the ionization energy distribution for a given ion charge. Many phenomena obey power law statistics, such as the Pareto distribution for instance \cite{Newman2005}. The latter implies that small occurrences are common, whereas large instances are rare. However, despite the fact that a power law models properly the tails of the empirical distribution, the largest events are significantly larger or smaller than what would be expected according to the power law. Such events are sometimes referred to as ``Dragon Kings'' as they indicate a departure from the generic process underlying the power law \cite{Pisarenko2012}. The extreme and/or rare events are difficult to understand and are often referred to as ``Black Swan'' phenomena \cite{Roccia2010,Janczura2012}. This means that they rarely happen and are unpredictable despite the fact that they have important consequences \cite{Wei2016}.

The Weibull distribution is the so-called ''extreme-value'' distribution. It successfully predicts the occurrence of extreme phenomena and rare events. It is particularly well-suited for data with heavy tails where values far from the maximum probability are still fairly common. The Weibull distribution is asymmetric, so that the probability of events before the mode is not the same as after. The distribution reads
\begin{equation*}
    f(x;\lambda ,\kappa)=\frac {\kappa}{\lambda }\left(\frac {x}{\lambda }\right)^{\kappa-1}e^{-(x/\lambda )^{\kappa}}
\end{equation*}
if $x\geq 0$ and 0 otherwise. $\kappa$ and $\lambda$ are respectively the strictly positive shape and scale parameters of the distribution. The two parameters can be determined by the knowledge of the mean value (or first-order moment):
\begin{equation*}
    \mu_1=\lambda\,\Gamma(1+1/\kappa)
\end{equation*}
and the variance:
\begin{equation*}
    v=\sigma^2=\mu_{2,c}=\mu_2-\mu_1^2=\lambda ^{2}\left[\Gamma(1+2/\kappa)-\Gamma(1+1/\kappa)^{2}\right],
\end{equation*}
where $\Gamma$ is the usual Gamma function. 
The skewness can be expressed as
\begin{equation*}
    \alpha_3={\frac {2\Gamma _{1}^{3}-3\Gamma _{1}\Gamma _{2}+\Gamma _{3}}{[\Gamma _{2}-\Gamma_{1}^{2}]^{3/2}}},
\end{equation*}
where $\Gamma_i=\Gamma(1+i/\kappa)$. It may also be written in terms of the mean value and the variance:
\begin{equation*}
    \alpha_3=\frac{\Gamma_3\lambda^3-3\mu_1\sigma^2-\mu_1^3}{\sigma^3}.
\end{equation*}
The excess kurtosis $-$kurtosis minus 3, where 3 is the value in the Gauss case$-$ is given by
\begin{equation*}
    \alpha_4-3={\frac {-6\Gamma _{1}^{4}+12\Gamma_{1}^{2}\Gamma _{2}-3\Gamma _{2}^{2}-4\Gamma _{1}\Gamma _{3}+\Gamma _{4}}{[\Gamma _{2}-\Gamma _{1}^{2}]^{2}}}.
\end{equation*}
It may also be written as
\begin{equation*}
    \alpha_4-3={\frac {\lambda ^{4}\Gamma_4-4\alpha_3\sigma ^{3}\mu -6\mu_1^{2}\sigma^{2}-\mu_1^{4}}{\sigma^{4}}}-3.
\end{equation*}
As mentioned above, the parameters $\kappa$ and $\lambda$ are determined by $\mu_1$ and $\sigma^2$. First, $\kappa$ is given by the equation
\begin{equation}
    \frac{\Gamma(1+2/\kappa)}{\Gamma(1+1/\kappa)^2}=\frac{\mu_2}{\mu_1^2}+1\label{eqk}
\end{equation}
and then $\lambda$ can be deduced from
\begin{equation*}
    \lambda=\frac{\mu_1}{\Gamma(1+1/\kappa)}.
\end{equation*}
Equation (\ref{eqk}) can be solved numerically, but one can also resort to an approximation. In fact,
\begin{equation*}
    \frac{\Gamma(1+2/\kappa)}{\Gamma(1+1/\kappa)^2}=\frac{(2/\kappa)\Gamma(2/\kappa)}{(1/\kappa)^2\Gamma(1/\kappa)^2}=2\kappa\frac{\Gamma(2/\kappa)}{\Gamma(1/\kappa)^2}.
\end{equation*}
It is possible to obtain a rather accurate approximant of $\kappa$ using the Pad\'e approximant of the Gamma function:
\begin{equation*}
    \Gamma(u)\approx \frac{1}{u}\frac{\left[12\gamma+(\pi^2-6\gamma^2)u\right]}{\left[12\gamma+(\pi^2+6\gamma^2)u\right]}.
\end{equation*}
The problem boils down to the resolution of the equation
\begin{equation*}
-\frac{\left[6\gamma(\gamma-\kappa)-\pi^2\right]\left[6\gamma(\gamma+2\kappa)+\pi^2\right]^2}{\left[6\gamma(\gamma-2\kappa)-\pi^2\right]^2\left[6\gamma(\gamma+\kappa)+\pi^2\right]}=\frac{\mu_2}{\mu_1^2}+1,
\end{equation*}
which can be recast, setting $\chi=\mu_2/\mu_1^2+1$, into
\begin{eqnarray*}
& &864\gamma^3(\chi-1)\kappa^3+288\gamma^3(\chi-1)\pi^2\kappa^2\nonumber\\
& &-[648(\chi-1)\gamma^5+72(\chi+1)\gamma^3\pi^2-30(\chi-1)\gamma\pi^4]\kappa\nonumber\\
& &+[6(\chi+1)\gamma^2-\pi^2(\chi-1)](36\gamma^4-\pi^4)=0
\end{eqnarray*}
and solved analytically using the Tartaglia-Cardano formulas \cite{McKelvey1984}. It is worth mentioning that an approximate solution of Eq. (\ref{eqk}) was proposed by Garcia \cite{Garcia1981}. If a rough approximation of $\kappa$ is sufficient, the following estimate may be useful. The Laurent series expansion of the Gamma function reads
\begin{equation*}
    \Gamma(z)\approx\frac{1}{z}-\gamma+\left(\gamma^2+\frac{\pi^2}{6}\right) z+O(z^2)
\end{equation*}
yielding, since $\Gamma(z+1)=z\Gamma(z)$:
\begin{equation*}
    \Gamma(z+1)\approx 1-\gamma z+\left(\gamma^2+\frac{\pi^2}{6}\right) z^2+O(z^3)
\end{equation*}
and, at second order in $z$:
\begin{equation*}
    \frac{\Gamma(1+2z)}{\Gamma^2(1+z)}\approx 1+\frac{\pi^2}{6}z^2.
\end{equation*}
Thus, using Eq. (\ref{eqk}), one gets, with $z=1/\kappa$:
\begin{equation}
    \kappa\approx\frac{\pi\mu_1}{\sigma\sqrt{6}}.\label{magic}
\end{equation}
Of course, the higher the value of $\kappa$, the higher the accuracy of approximant (Eq. \ref{magic}), as can be checked in Table \ref{table_magic}.

\begin{table}[!ht]
    \centering
    \begin{tabular}{|c|c|}\hline
        Exact value of $\kappa$ & Equation (\ref{magic})\\\hline
         3 & 3.529 \\\hline
         5 & 5.599 \\\hline
         10 & 10.660 \\\hline
         30 & 30.706 \\\hline
         50 & 50.716 \\\hline
         70 & 70.720 \\\hline
         100 & 100.723 \\\hline
    \end{tabular}
    \caption{Illustration of the accuracy of Eq. (\ref{magic}) for different values of the Weibull parameter $\kappa$.}
    \label{table_magic}
\end{table}

\subsubsection{Comparisons and discussion}

\begin{figure}[!ht]
\begin{center}
\includegraphics[scale=0.5]{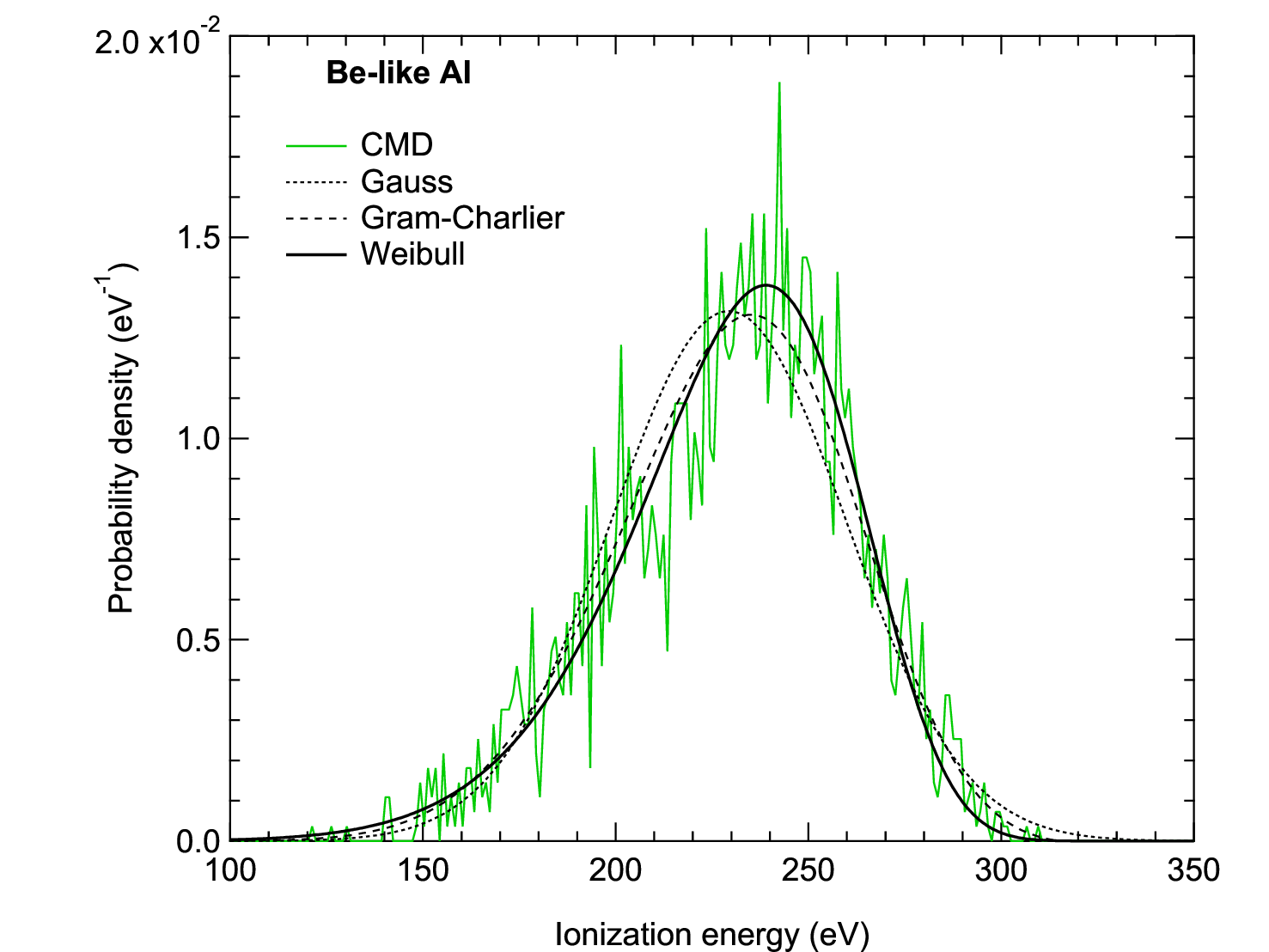}
\end{center}
\caption{Normalized distributions of the ionization energy of Be-like aluminum. Density=2.7 g/cm$^3$, temperatures: $kT_e=$50 eV and $kT_i=$300 K.}
\label{Distrib_BeAl}
\end{figure}

Figure \ref{Distrib_BeAl} shows the CMD ionization energy distribution of Be-like aluminum. The ionic density and the temperatures ($T_e$: electron, $T_i$: ion) are inferred from the experiment \cite{Ciricosta2016}. The average energy and the moments of this distribution provide the appropriate Gauss, Gram-Charlier and Weibull distributions. The CMD distribution shows noise that can be important as we will see below. We notice a small asymmetry in the Gram-Charlier distribution, which is consistent with the small deviation of the skewness from the value $\alpha_3=0$ corresponding to the Gauss distribution (see Fig. \ref{alpha} below). The Weibull distribution shows the best agreement with the CMD one. Nevertheless, significant noise still remains. The ionization energy used in the analytical distributions is taken to be the average over the CMD one.

\begin{figure}[!ht]
\begin{center}
\includegraphics[scale=0.5]{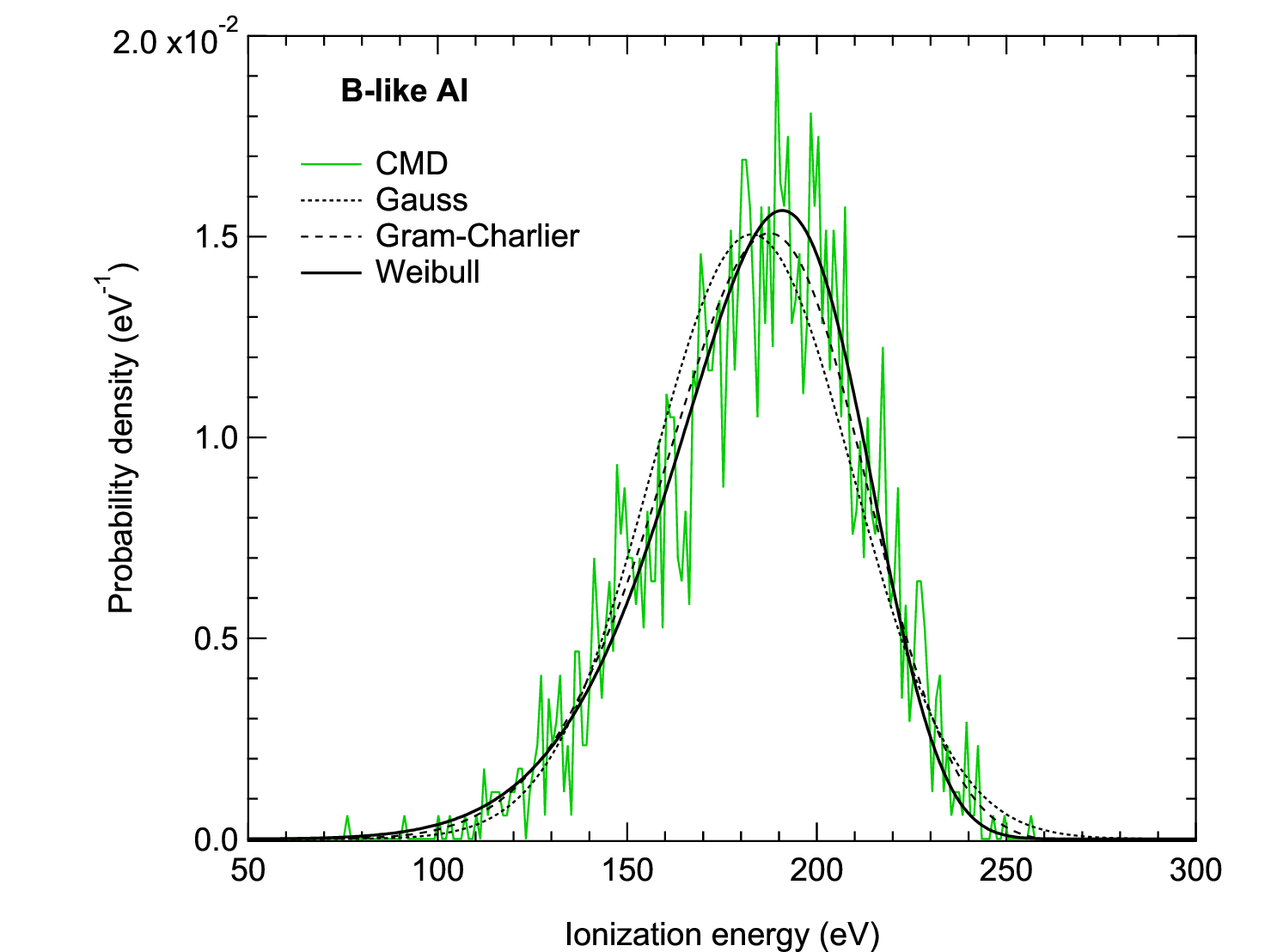}
\end{center}
\caption{Same as Fig. \ref{Distrib_BeAl} for B-like aluminum.}
\label{Distrib_BAl}
\end{figure}

Figure \ref{Distrib_BAl} represents the same distributions for B-like aluminum. As in the previous case, the Weibull distribution shows the best agreement with the CMD one.

The fact that the Weibull distribution provides a better depiction of the CMD distribution, in particular of its asymmetry and peakedness, is consistent with the fact that the ionization/recombination processes taken into account in the simulation are rare events.

\subsubsection{Statistical properties}

\begin{figure}
\centering
\includegraphics[scale=0.45]{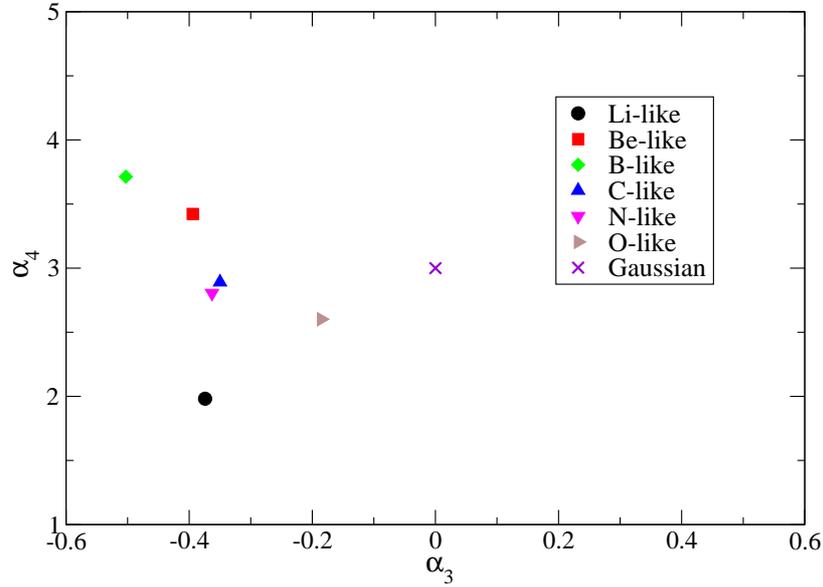}
\caption{Skewness ($\alpha_3$) and kurtosis ($\alpha_4$) of aluminum ions. The cross mark ($\alpha_3,\alpha_4)$=(0,3) corresponds to the Gauss distribution.}\label{alpha}
\end{figure}

\begin{figure}
\centering
\includegraphics[scale=0.45]{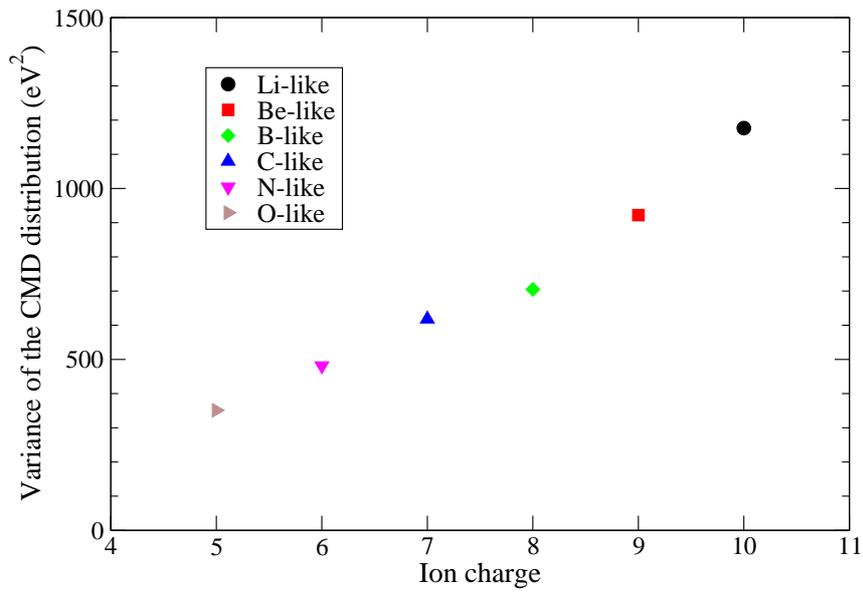}
\caption{Variances of aluminum ions. Density and temperatures, as Fig. \ref{Distrib_BeAl}.}\label{variance}
\end{figure}

Figure \ref{alpha} represents the pairs $(\alpha_3,\alpha_4)$ of the CMD distribution of O- to Li-like aluminum. We can see that all the distributions have a negative skewness, which means that the tail of the left side of the distribution is longer than the tail of the right one. The mean and median $-$the ``middle'' value in a list of numbers$-$ will be less than the mode, \textit{i.e.}, the value that occurs most often. The Gaussian, which is symmetrical, has a skewness equal to zero. On the other hand, the kurtosis is either larger than 3 (the Gaussian value) or smaller. The distribution is said leptokurtic in the first case and platykurtic in the second one. The CMD distribution is leptokurtic for B- and Be-like ions, and platykurtic for O-, N-, C-like ions. The Li-like case has a kurtosis close to 4/5, which means that is resembles a rectangular function, and for C-like and N-like ions, the distribution is almost mezokurtic (the kurtosis is very close to 3).
Figure \ref{variance} shows that the variance of the CMD distribution increases with the ion charge. We notice that the higher the ion charge, the more important the spread of the IPD around its average value. This may be explained by the depth of the potential well rather than by the number of bound electrons, since in the simulation the only bound electron is the one which is ionized.

\section{Noise reduction}\label{Noise Reduction}

In Ref. \cite{Benredjem2022}, the ionization energy was defined by an average over the CMD distribution. Unfortunately, this definition involves the whole distribution, including the numerical noise that is inherent to the CMD method. Moreover, the plasma fluctuations due to ion dynamics are partially hidden by this numerical noise. It is crucial to eliminate the noise, in order to obtain a new distribution which represents the actual plasma fluctuations. In practice, we then obtain a new distribution composed of a small set of Gaussian peaks. 

A ``peak'' is defined as a local maximum in the magnitude of a signal. Peak searching algorithms detect either a single or multiple peaks which magnitude is larger than a specified threshold and adjust the range to obtain the finest tuning as possible. Usually, the peaks are detected in a noisy signal, and the only practical constraints to be made in the peak search are to have a range and magnitude threshold. Peak search methods vary by algorithm implementation such as correlation, filtering, demodulation, sliding average, etc. techniques \cite{Igorpro}. \\
\indent Peak detection algorithm often applies wavelet transform \cite{Meyer1992,Ventzas2011}, which is a broadly applicable analysis tool that is analogous to the more familiar Fourier transform. Fourier analysis is commonly used to express spectral data in terms of frequency components and associated phases. While transformation into this frequency-phase space is useful for many data analysis practices, the most intuitive transformation space for peak identification is a peak width-position space. A wavelet series is a representation of a square-integrable function by a certain orthonormal series generated by a wavelet, \textit{i.e.}, a brief wave-like oscillation with an amplitude that begins at zero, increases or decreases, and then returns to zero one or more times. Wavelets have specific properties that make them useful for signal processing \cite{Young1993}. In the formalism of wavelet analysis, this type of transform can be built by an appropriate choice of a mother wavelet such that peak width and position are accessed through dilation and translation of this wavelet \cite{Gregoire2011}. Many multipeak-finding softwares rely, at least in a partial way, on wavelet transform \cite{Igorpro}. \\
\indent We have compared the Gaussian multipeaks fitting based on a denoising technique with a continuous wavelet transform of the signal. The results are rather close, but we have chosen the former because the Gaussian representation allows for the derivation of analytical formulas for the EII cross section (see section \ref{New expression}).
In practice, the net signal, \textit{i.e}, the set of the Gaussian peaks, is the difference between the -crude- CMD ionization potential distribution and the noise. As a result, one obtains a set of ionization energies for each ion charge. We assume that the relevant (most probable) ionization energy is represented by the Gauss peak with the highest signal-to-noise ratio (SNR). The useful parameters of each Gaussian peak are its position, height and signal-to-noise ratio. As will be shown below, the width does not matter in cross section calculations, as long as one deals with a Gaussian shape.

\begin{figure}[ht]
\begin{center}
\includegraphics[scale=0.55]{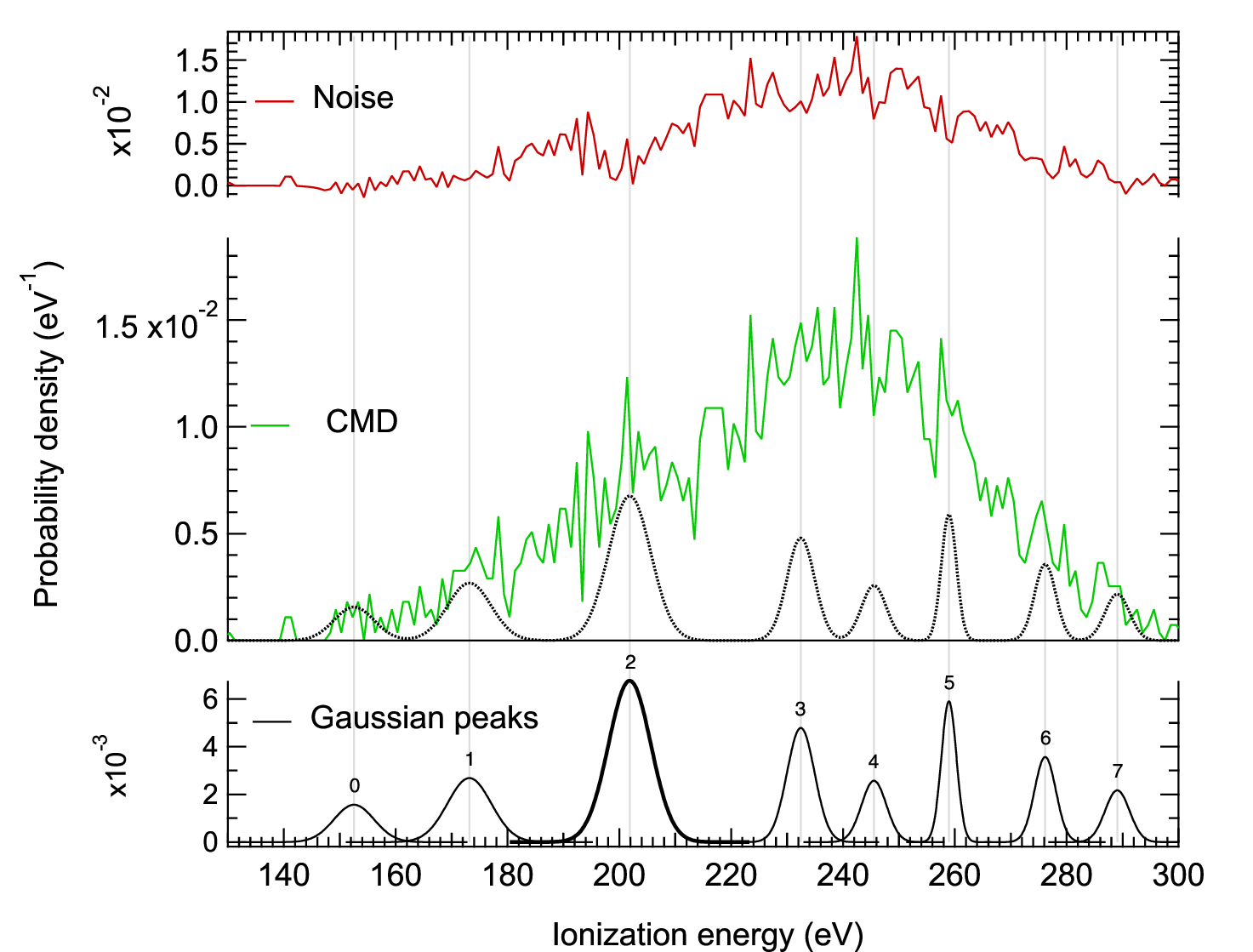}
\end{center}
\caption{Normalized distribution of the ionization potential of Be-like Al: CMD, Noise and the set of Gaussian peaks. Density and temperatures, as in Fig. \ref{Distrib_BeAl}. The set of Gaussian peaks (black curve) is the difference between the CMD distribution (green curve) and the noise (red curve).}
\label{Multipeak_Be-like_Al}
\end{figure}

Let us focus on Be-like aluminum. In Fig. \ref{Multipeak_Be-like_Al}, we show the distribution of the ionization energy. After extracting the noise from the original CMD distribution, we are left with a set of eight Gaussian peaks. We assume that the most probable ionization energy is represented by the highest peak (bold curve, \#2), which will be named Principal Gaussian Peak (PGP). The corresponding energy $E_{i2}=201.85$ eV (see Table \ref{Tab:IPD}), which is substantially lower than $E_{i1}$, the value obtained in our previous work. Knowing the ionization energy of the isolated ion (399.37 eV), we obtain an ionization potential depression IPD$_2=197.52$ eV (see Table \ref{Tab:IPD}). This value shows a better agreement with the experiment \cite{Ciricosta2016} than all other calculations.

Table \ref{Tab:IPD} presents the ionization energy of the isolated ($E_{i0}$) and the surrounded ($E_{i1}$ or $E_{i2}$) ions and the ionization potential depression (IPD$_{1}$ or IPD$_{2}$). Index "1" refers to an average of the ionization energy over the whole CMD distribution while index "2" corresponds to an average over the PGP. In the last row (O-like aluminum, $z=5$) we have two ionization energies and then two IPD values because after eliminating the noise from the CMD distribution we get two Gaussian peaks with the same peak value.

\begin{table}
\caption{Ionization energy and IPD (in eV). $E_{i0}$: ionization energy of the isolated ion; $E_{i1}$: average over the CMD distribution, $E_{i2}$: average over Principal Gaussian Peak, IPD$_1=E_{i1}-E_{i0}$, IPD$_2=E_{i2}-E_{i0}$.}
\begin{tabular}{|l|l|l|l|l|l|}
\hline
Ion & $E_{i0}$ & $E_{i1}$ & $E_{i2}$ & IPD$_1$ & IPD$_2$\\
\hline
Be-like & 399.37 & 229.33 & 201.85& 170.04 & 197.52\\
B-like & 330.11 & 182.85 & 147.24& 147.26 & 182.87\\
C-like & 284.60 & 158.01 & 146.24& 126.58 & 138.36\\
N-like & 241.44 & 130.92 & 130.10& 110.52 & 111.34\\
O-like & 190.48 & 88.58 & 80.16\ (89.18)& 101.90 & 110.32\ (101.3)\\
\hline
\end{tabular}\label{Tab:IPD}
\end{table}

\begin{figure}[ht]
\begin{center}
\includegraphics[scale=0.55]{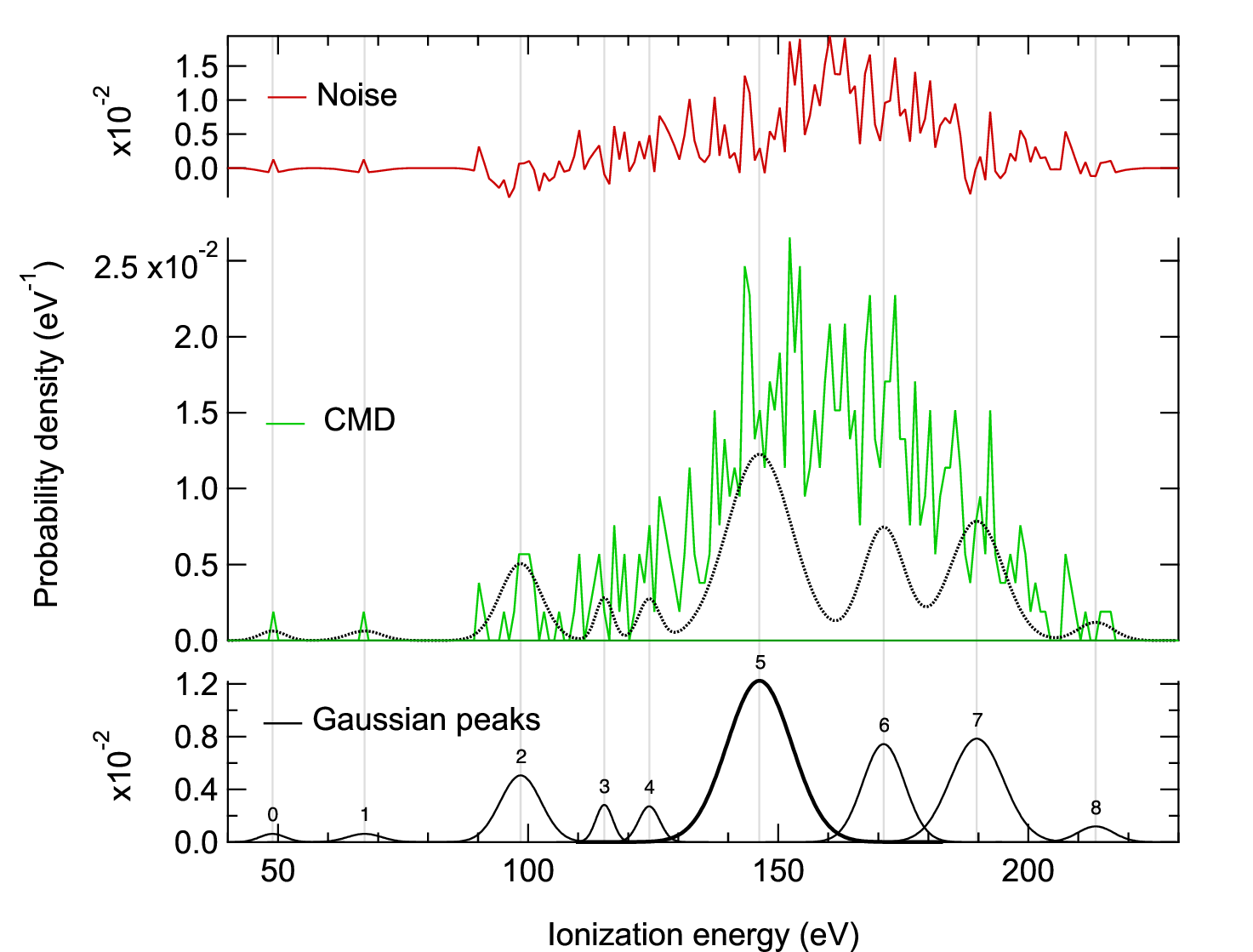}
\end{center}
\caption{Same as Fig. \ref{Multipeak_Be-like_Al} for C-like aluminum. The set of Gaussian peaks (black curve) is the difference between the CMD distribution (green curve) and the noise (red curve).}
\label{Multipeak_C-like_Al}
\end{figure}

Figure \ref{Multipeak_C-like_Al} shows the same distribution as above for C-like aluminum. Here, the CMD distribution is represented by 9 Gaussian peaks, the highest one being centered on 146.24 eV (bold curve, \#5). Here again, we assume that the most probable ionization energy is associated with this peak. In this case, the ionization energy of the isolated ion is 284.60 eV. The obtained IPD is then equal to 138.36 eV, a value to be compared to our previous result \cite{Benredjem2022}: 126.58 eV. As in the Be-like case, the present calculation is in better agreement with the experiment \cite{Ciricosta2016} than our previous one.

\begin{figure}[ht]
\begin{center}
\includegraphics[scale=0.5]{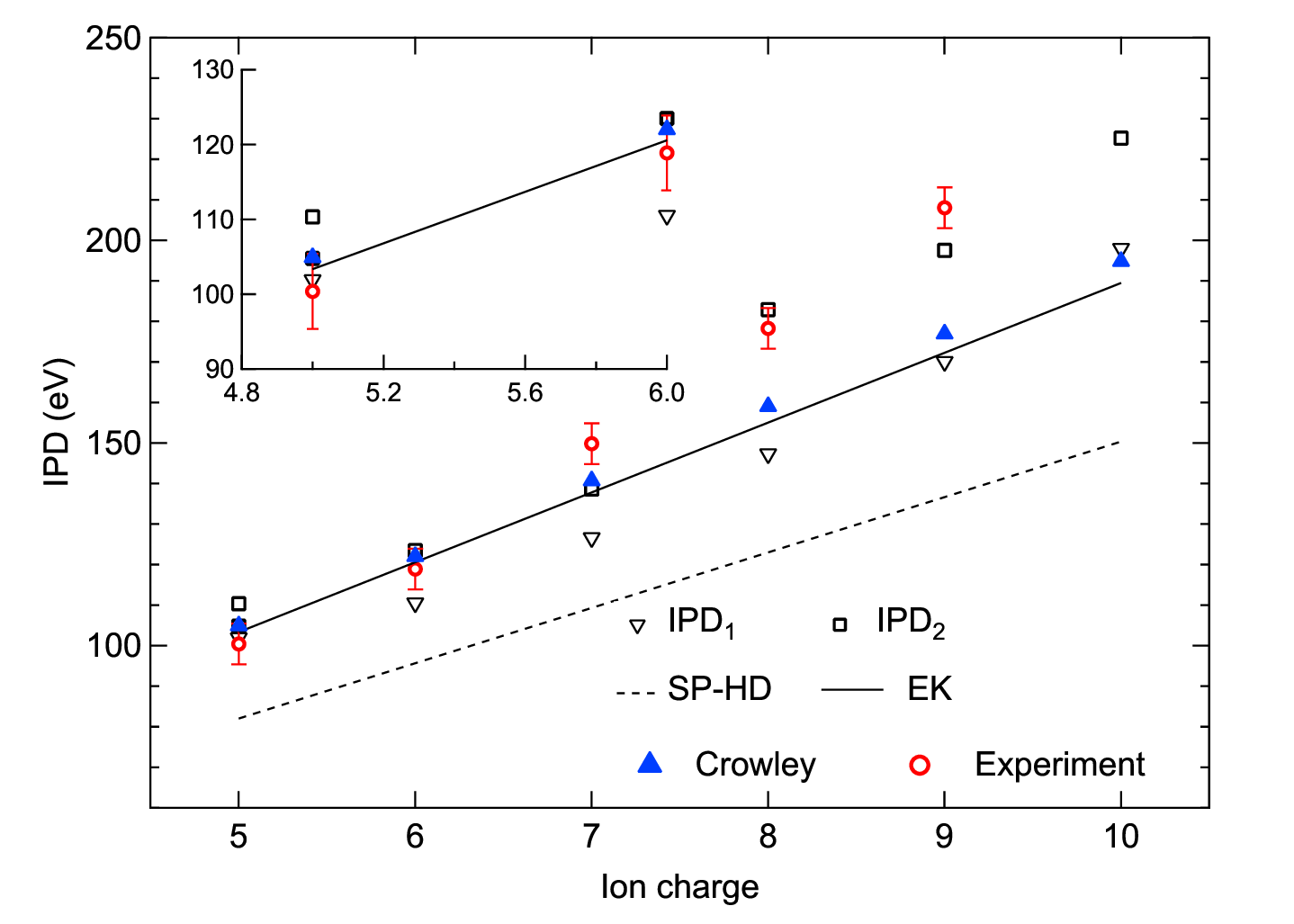}
\end{center}
\caption{IPD as a function of the ion charge in aluminum. IPD$_1$: average over the CMD distribution (see Table \ref{Tab:IPD}), IPD$_2$: average over the Principal Gaussian Peak (see Table \ref{Tab:IPD}), Experiment: Ref. \cite{Ciricosta2016}, Crowley: Ref. \cite{Crowley2014}, 
SP-HD: Stewart-Pyatt (high-density limit), EK: Ecker-Kr\"{o}ll (with $C=1$). Density and temperatures, as in Fig. \ref{Distrib_BeAl}.}
\label{IPDcomparison}
\end{figure}

The IPD of O- to Li-like ions is represented in Fig. \ref{IPDcomparison}. The comparison with experimental results shows that:\\ (i) Stewart-Pyatt formula largely underestimates the IPD.\\(ii) Ecker-Kr\"{o}ll formula, with $C=1$, is satisfactory for O- and N-like ions but shows large discrepancies for higher ion charges. In fact, this trend was noticed in recent calculations on silicon for similar densities \cite{Zeng2020}.\\(iii) Crowley's results \cite{Crowley2014} show a good agreement for O- to C-like ions.\\ (iv) Our new results (IPD$_2$) show\\
\indent $-$ the same level of agreement with experiment as those of Crowley for O- to C-like,\\
\indent $-$ a better agreement with experiment than our previous $-$IPD$_1-$ results,\\
\indent $-$ a better agreement with experiment than all other calculations, for the highest ion charges.

In the following, we concentrate on the EII cross sections and on how they are affected by density effects in plasmas.

\section{EII cross section}\label{EII cross section}

\subsection{New expression}\label{New expression}

The cross section $\sigma(E|E_i)$ varies with the incident electron energy $E$ and is a function of the ionization energy $E_i$ of the target ion. In our previous work \cite{Benredjem2022}, we used the Lotz formula \cite{Lotz1967} with an ionization energy averaged over the CMD distribution. With the aim of accounting for the plasma fluctuations, we now propose to average the cross section over an ionization energy distribution $P(E_i)$ (CMD or another one). The new cross section $\sigma(E)$ is then written as
\begin{equation}
    \sigma(E)=\langle\sigma(E|E_i)\rangle_{P}=\int \sigma(E|E_i) P(E_i) dEi,\label{New-definition}
\end{equation}
where $P$ is a chosen distribution.

\begin{figure}[ht]
\begin{center}
\includegraphics[scale=0.5]{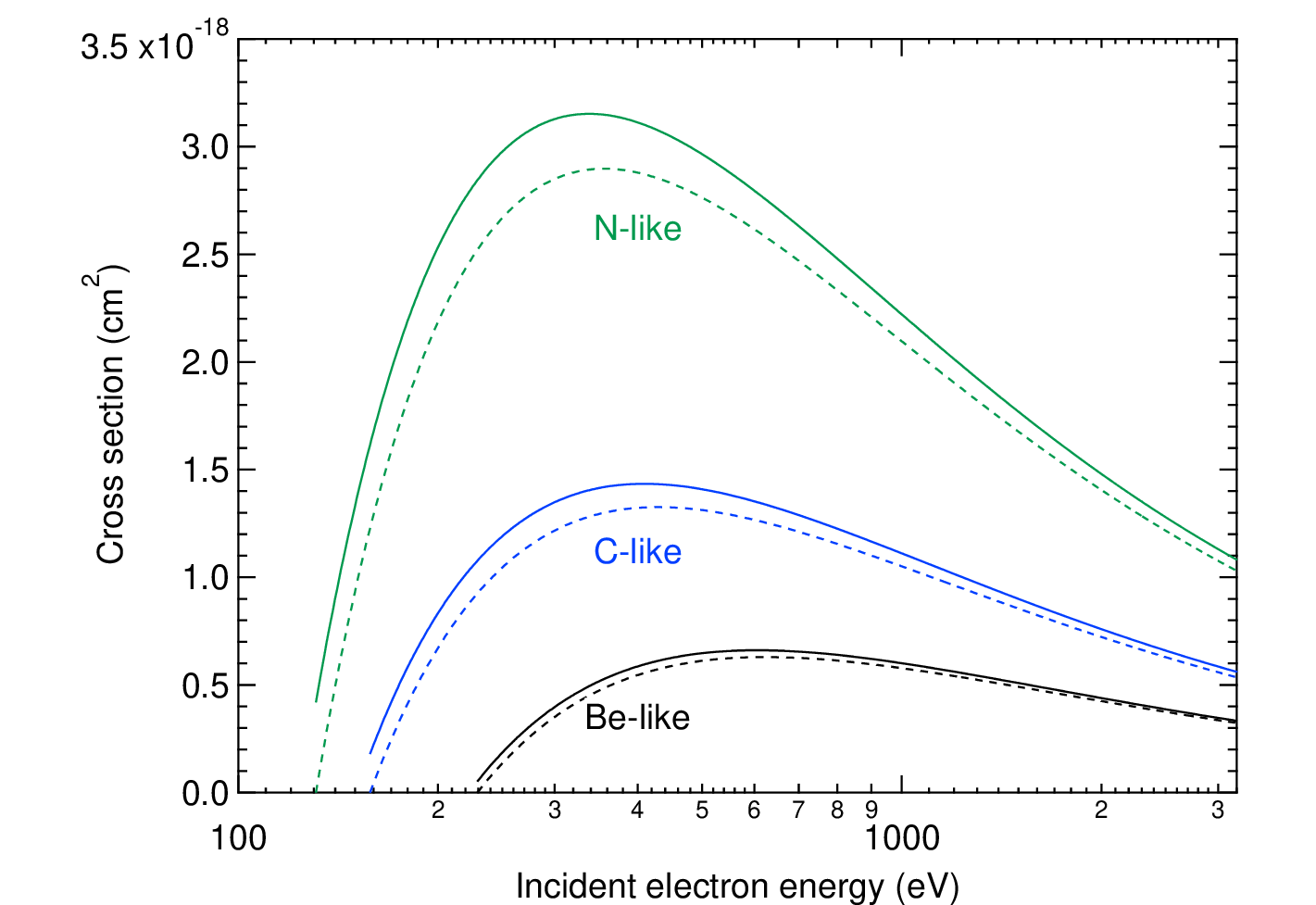}
\end{center}
\caption{Cross section of Be-, C- and N-like aluminum. Comparison of the two methods. Eq. (\ref{XS-Lotz}): dashed curve, Eq. (\ref{New-definition}): full curve. Plasma conditions, as in Fig. \ref{Distrib_BeAl}.}
\label{X1XS2Lotz}
\end{figure}

Let us first consider the Lotz formula:
\begin{equation}
    \sigma(E|E_i)=A\, \xi\,\frac{\ln(E/E_i)}{E\,E_i},\label{XS-Lotz}
\end{equation} 
where $\xi$ is the number of electrons in the subshell from which the ionization occurs, and $A=4.5\times 10^{-14}$ cm$^{2}\cdot$ eV$^{2}$, if the energies are in eV and the cross section in cm$^{2}$. To investigate the effect of the fluctuations we consider the CMD distribution and compare $\sigma(E|\bar{E_i})$ (see Eq. (\ref{XS-Lotz})), where $\bar{E_i}$ is the average ionization energy, and the average cross section $\sigma(E)$ (see Eq. (\ref{New-definition})). The comparison (see Fig. \ref{X1XS2Lotz}) clearly shows that the fluctuations play a significant role. However this effect decreases for increasing ion charges.

\subsection{Effect of the plasma fluctuations on the cross sections}
\begin{figure}[ht]
\begin{center}
\includegraphics[scale=0.5]{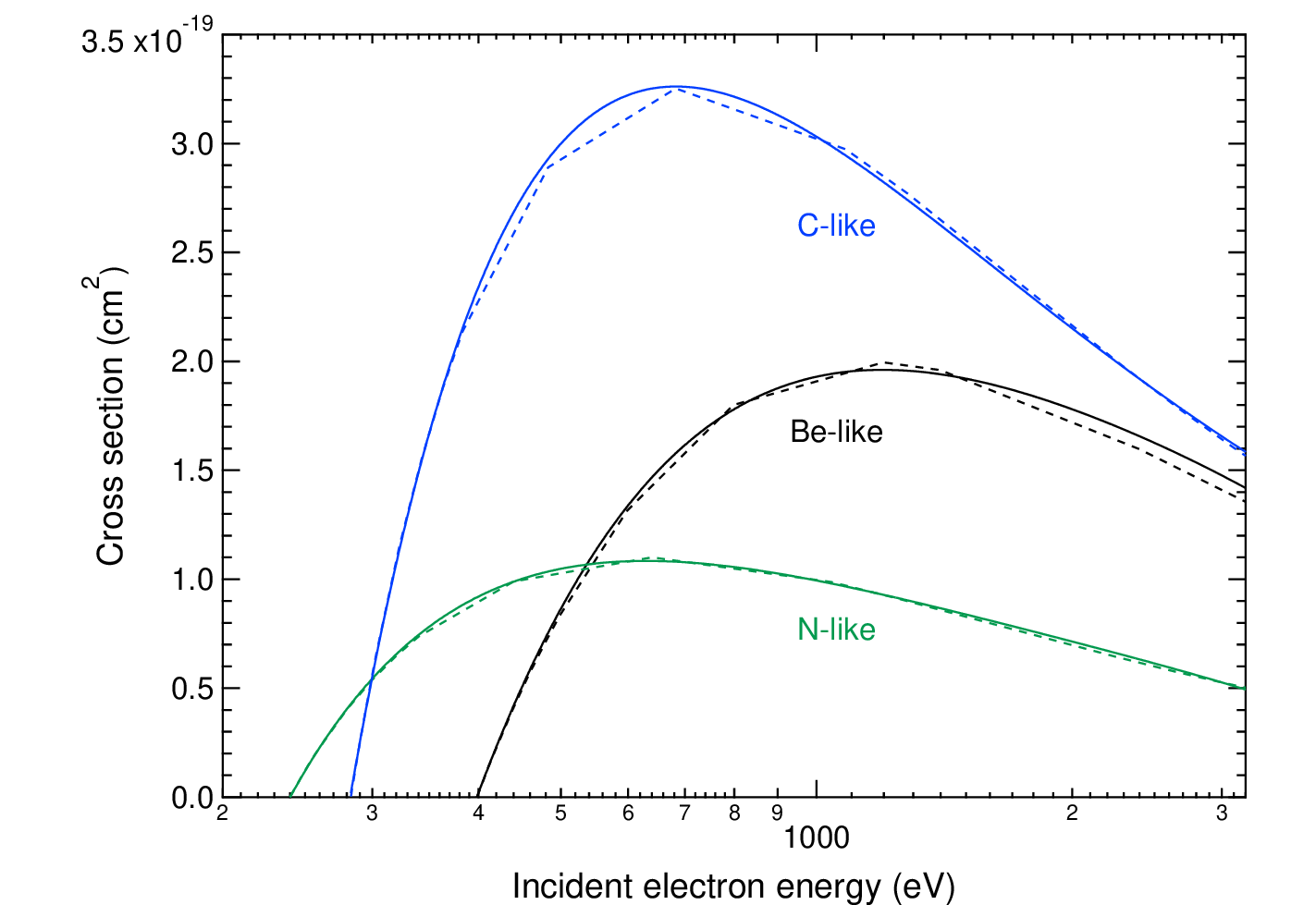}
\end{center}
\caption{Cross sections of isolated Be-, C- and N-like ions. Dashed curves: FAC results, Full curves: fit of FAC cross sections with formula in Eq. (\ref{XS-Kim}) and adjustable parameters in Table \ref{Fitting parameters}.}
\label{XSFfitKBeNClike}
\end{figure}

\begin{figure}[ht]
\begin{center}
\includegraphics[scale=.5]{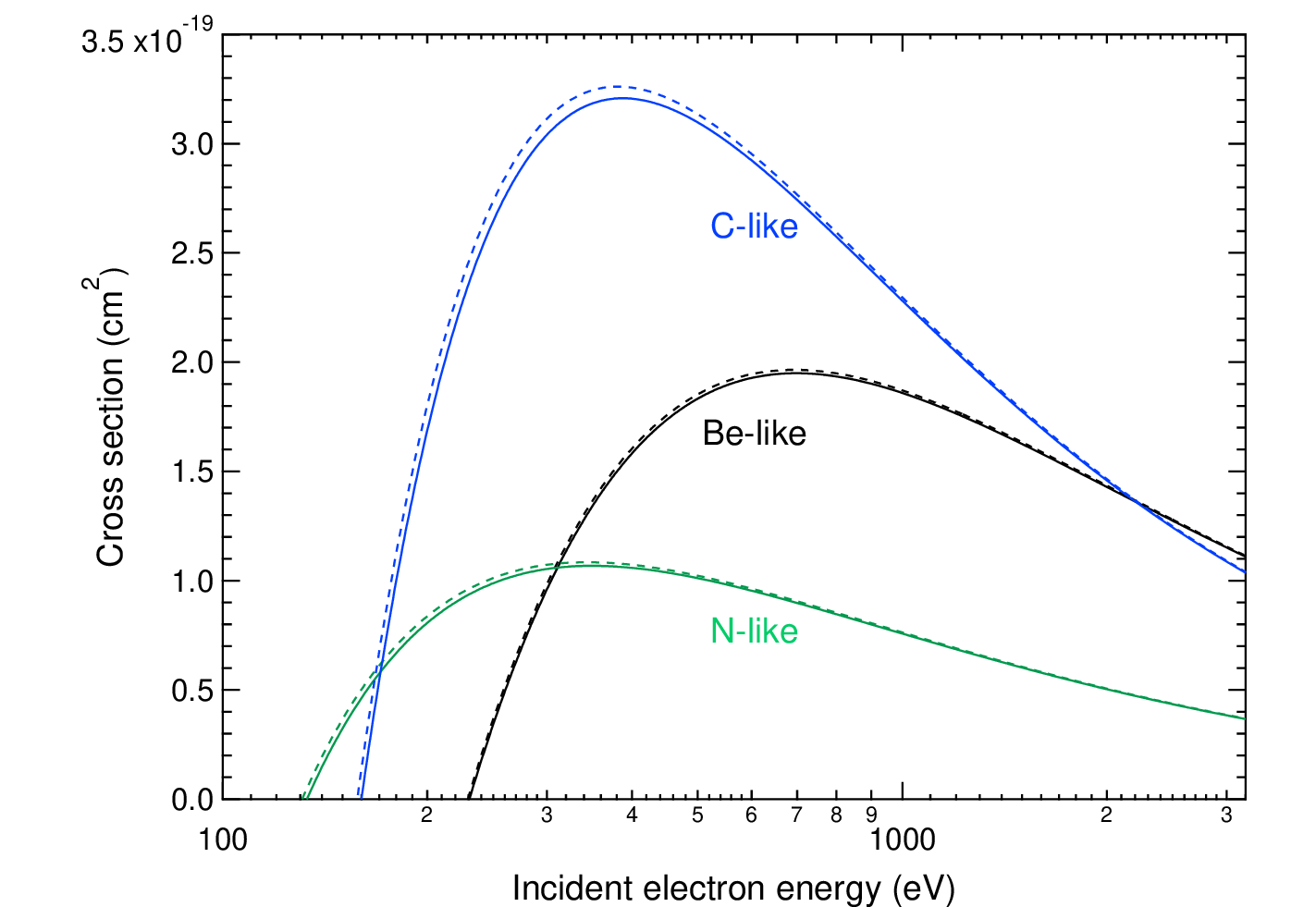}
\end{center}
\caption{Cross section of Be-, C- and N-like aluminum. Dashed curves: $E_i=E_{i,1}$, Full curves: Eq. (\ref{New-definition}) with the CMD distribution. Density and temperatures, as in Fig. \ref{Distrib_BeAl}.}\label{XS1K_2K_G_GC_BeNClike}
\end{figure}

In this work, we rather consider the cross section:
\begin{equation}
    \sigma(E|E_i)=\sigma(x)=B_0 \frac{\ln(x)}{x}+\sum_{l=1}^N\frac{B_l}{x^l},\label{XS-Kim}
\end{equation}
where $x=E/E_i$. This formula is an extension of the Kim cross section \cite{Kim1992} to higher powers of $x$. $B_0$ and $B_l$ are adjustable parameters. By varying these parameters we obtain a very good fit of accurate cross sections of isolated ions. The reference cross sections are calculated by the FAC code \cite{Gu2008} for a set of incident electron energies. Kim's formula was limited to $N=2$. Considering isolated aluminum ions and focusing on the ionization from ground state, it was necessary to add a third term into the expansion in order to obtain a good fit of the accurate cross sections provided by FAC \cite{Gu2008} or HULLAC \cite{BarShalom2001} codes. In Fig. \ref{XSFfitKBeNClike}, we compare the cross section calculated by FAC to the cross section given by Eq. (\ref{XS-Kim}) for C-, Be- and N-like isolated ions. The fitting procedure is satisfactory. The values of the parameters $B_l\ (l=0-3$) are given in Table \ref{Fitting parameters}. As in Sec. \ref{New expression}, we investigate the effect of the plasma on the cross section. Knowing the values of the above parameters, we calculate the cross section by taking into account the fluctuations and the continuum lowering. In a first calculation, the ionization energy is taken to be the average over the CMD distribution, $E_{i,1}$, and the cross section is given by Eq. (\ref{XS-Kim}), where $E_i=E_{i,1}$. In the second one, we average the cross section over the CMD distribution. Figure \ref{XS1K_2K_G_GC_BeNClike} shows the Be-, C- and N-like aluminum cross sections obtained by the two methods. We can see that the difference is negligible, in contradiction with the previous calculation relying on the Lotz formula. It is worth mentioning that weighting over the CMD, Gaussian, Gram-Charlier or Weibull distributions yields very close cross sections.

\begin{table}
\caption{\label{Fitting parameters}Values (in $10^{-19}$ cm$^{2}$) of the adjustable parameters of C-, N- and Be-like aluminum.}
\centering
\begin{tabular}{|l|c|c|c|c|}
\hline
 & $B_0$ & $B_1$ & $B_2$ & $B_3$ \\
 \hline
Be-like &3.80 & 3.89 & -7.63 & 3.76 \\
C-like &4.79 &7.09& -9.18 & 2.10 \\
N-like & 1.87 & 2.21 & -3.41 & 1.21 \\
\hline
\end{tabular}\\
\end{table}

In Sec. \ref{Noise Reduction}, we described a method which allows us to denoise the CMD distribution. This method provided new IPDs which are in better agreement with experiment than our previous calculations \cite{Benredjem2022}. In particular, we identified, for each ion stage, a Gaussian peak that represents the ionization energy distribution. In the following, we calculate the cross section averaged over this new distribution.

\subsection{Cross section involving the PGP}
The cross section is given by Eq. (\ref{XS-Kim}) and Eq. (\ref{New-definition}) where $P$ is the PGP distribution. As said above, the PGP is selected by comparing the SNR values of the highest peaks. The cross section then reads:
\begin{equation*}
    \sigma(E)=\frac{1}{\sqrt{2\pi\,v}}\int_0^{\infty}\sigma(E|E_i)\exp\left[-(E_i-\mu_1)^2/(2v)\right]dE_i.
\end{equation*}
It can be expressed in the compact form:
\begin{equation*}
    \sigma(E)=B_0\left[I_1\,\ln E-K\right]+\sum_{l=1}^3 B_l\,I_l,
\end{equation*}
where 
\begin{equation}
    I_l=\frac{1}{\sqrt{2\pi\,v}}\frac{1}{E^l}\int_0^{\infty}E_i^l\exp\left[-(E_i-\mu_1)^2/(2v)\right]dE_i\label{eqn:Il} 
\end{equation}
and 
\begin{equation}
    K=\frac{1}{\sqrt{2\pi\,v}}\frac{1}{E}\int_0^{\infty}E_i\ln(E_i)\exp\left[-(E_i-\mu_1)^2/(2v)\right]dE_i. \label{eqn:K}
\end{equation}
The calculation of $I_l$ and $K$ is presented in Appendix \ref{Integrals}. Since the results involve special functions, it may be interesting to derive alternative analytical expressions based on the representation of the Gaussian by cubic splines. The cross section calculated by this method is given in Appendix \ref{Cubic_splines}.

\begin{figure}[ht]
\begin{center}
\includegraphics[scale=.5]{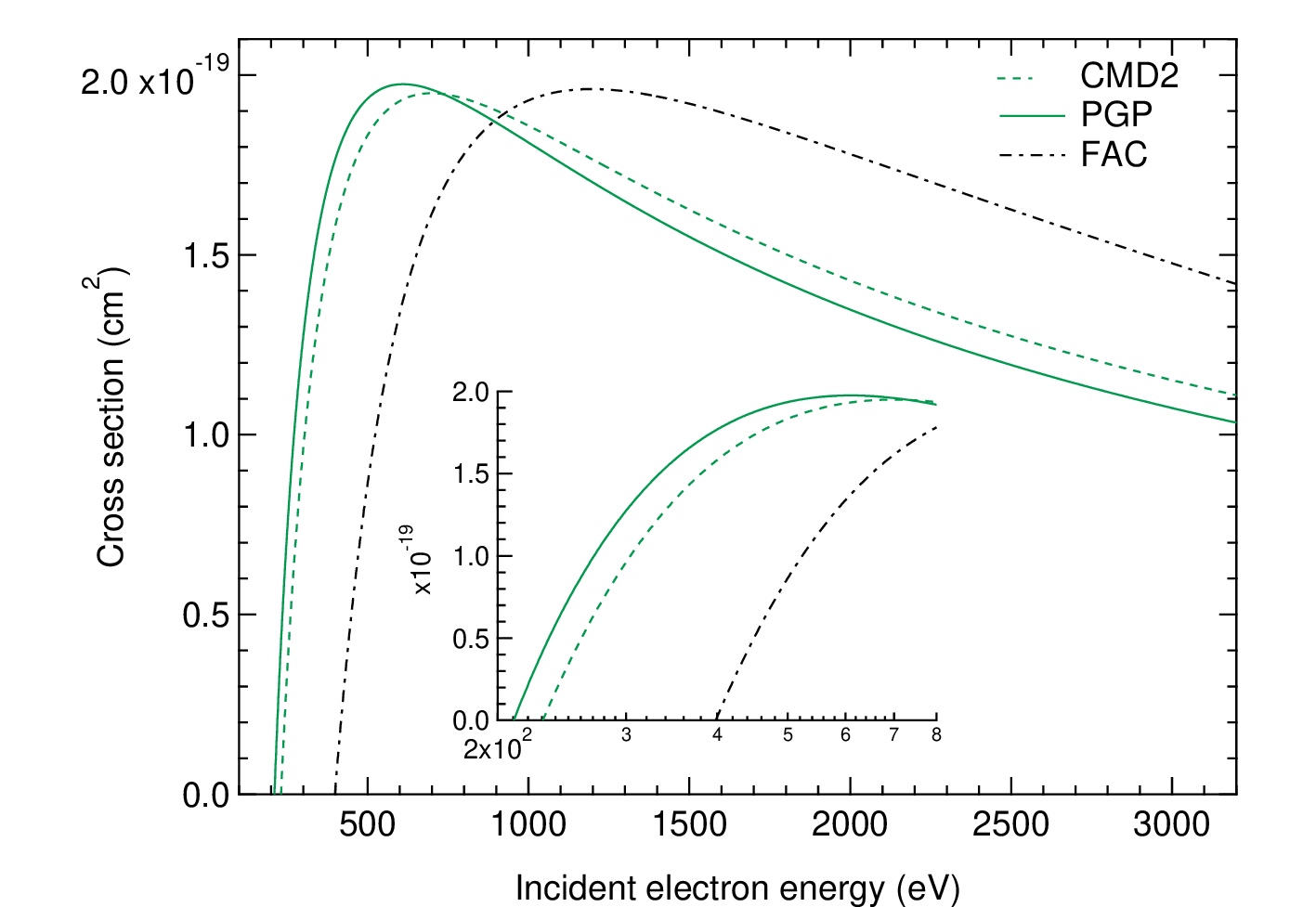}
\end{center}
\caption{Cross section of Be-like Al. Weighting by the CMD distribution: CMD2, by the principal Gaussian peak: PGP, dashed line: FAC results.}\label{XS2K_GNRK_Belike}
\end{figure}

% ---------figure Carbon-like------------
\begin{figure}[h!]
\begin{center}
\includegraphics[scale=.5]{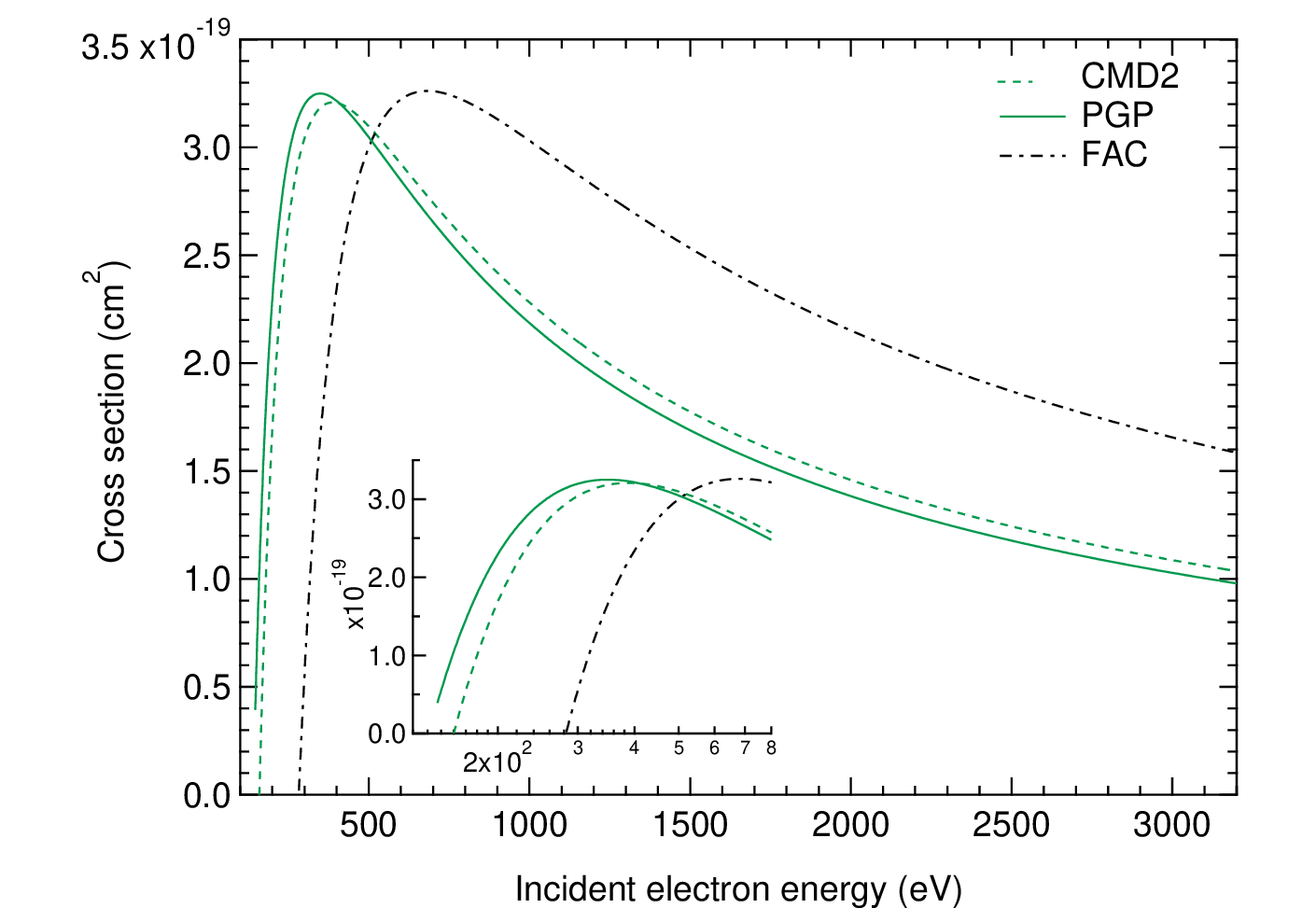}
\end{center}
\caption{Same as Fig. \ref{XS2K_GNRK_Belike} for C-like Al.}\label{XS2K_GNRK_Clike}
\end{figure}

% ---------fin Carbon-like------------
\begin{figure}[h!]
\begin{center}
\includegraphics[scale=0.5]{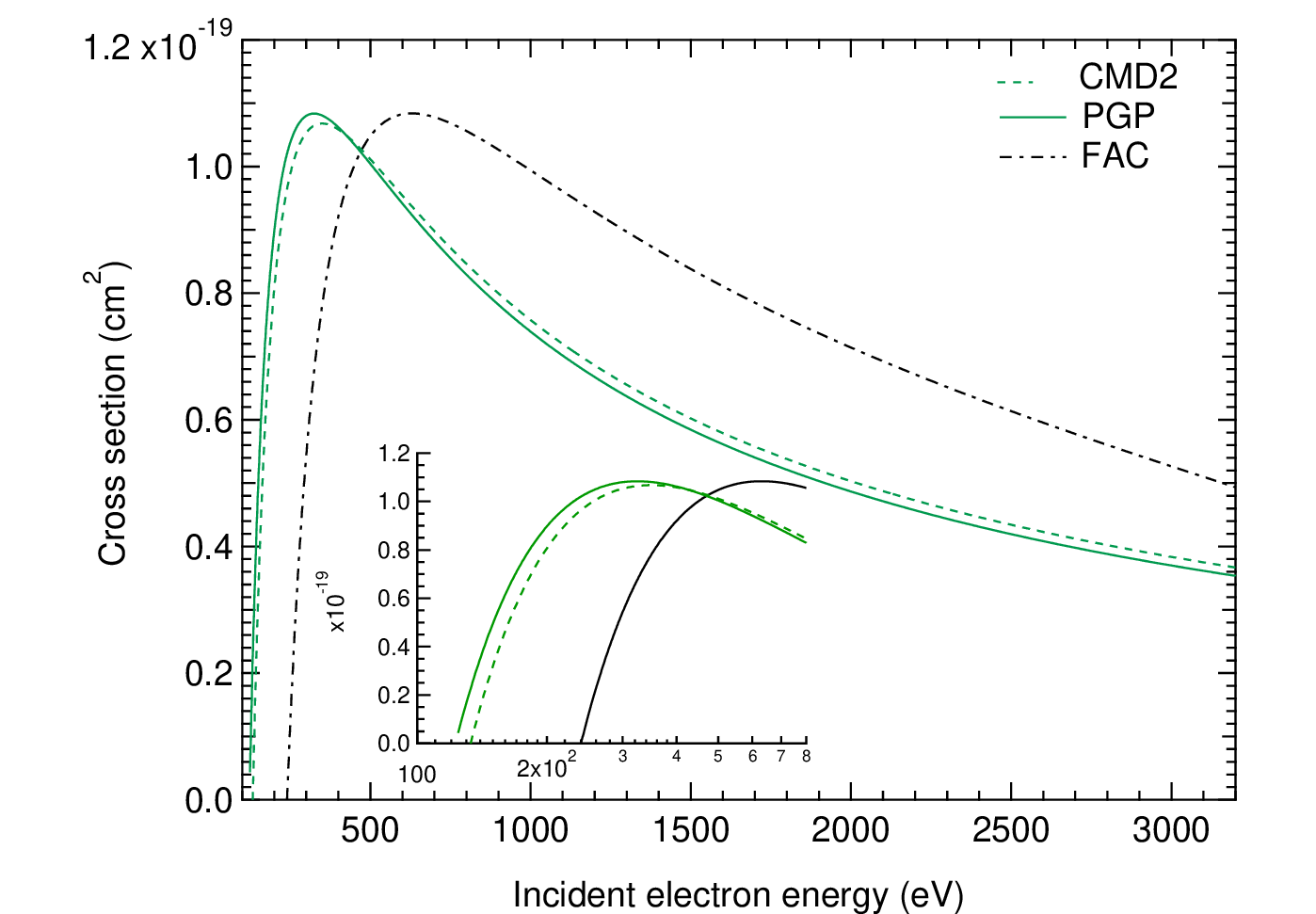}
\end{center}
\caption{Same as Fig. \ref{XS2K_GNRK_Belike} for N-like aluminum.}
\label{XS2K_GNRK_Nlike}
\end{figure}

Figures (\ref{XS2K_GNRK_Belike})--(\ref{XS2K_GNRK_Nlike}) display the Be-, C- and N-like cross sections, respectively. The two calculations (analytical and cubic\textcolor{green}{-}splines method) yield identical results. The black curve is the cross section of the isolated ion, calculated with FAC. The full green curve shows the cross section $\sigma_2(E)$ which is obtained by averaging the ionization energy over the PGP distribution. The dashed curve, $\sigma_1(E)$, is obtained by averaging over the whole CMD distribution. We can conclude that
(i) The comparison with the isolated ion cross section shows that the continuum lowering plays an important role.
(i) Averaging over the CMD or the PGP distribution yield\textcolor{green}{s} a difference that increases with increasing ion charge. In fact, the relative difference between the corresponding IPDs (see Table \ref{Tab:IPD}) increases with the ion charge.

\section{Conclusion and short-term prospective}
We studied the ionization of aluminum by electron impacts in a high-density plasma where fluctuations due to ion dynamics are present. The classical molecular dynamics method is appropriate to account for such fluctuations. It provides a distribution of the ionization energy. We showed that this distribution can be reproduced by a small set of Gaussian peaks, each one representing an ionization energy. By eliminating the noise from the distribution, we were able to select one Gaussian, the one that is characterized by the highest signal-to-noise ratio. For all ion charges, from O- to Be-like, the ionization potential depression associated to this Gaussian is in better agreement with experiment than all other calculations.\\
\indent The second part of our work was devoted to the effects of the continuum lowering and the plasma fluctuations on the cross section. We proposed a new expression of the cross section. The fluctuation effect is very small when one uses a cross-section formula similar to Kim's one \cite{Kim1992}. On the contrary, the Lotz formula \cite{Lotz1967} is more sensitive to plasma fluctuations. In both cases, the lowering of the ionization potential affects the cross section, in particular the ionization threshold.\\
\indent Further developments within classical molecular dynamics are welcome. In fact, it is important to improve the statistics, by increasing the numbers of particles and time intervals. This requires larger simulation ``boxes'' and consequently time consuming calculations. As a result, the noise would be reduced giving then a more convenient smaller set of Gaussian peaks. Also, considering the quasi-bound states could give a new insight in the definition of the ionization energy.\\
\indent We plan to calculate the IPD of silicon and magnesium, for which experimental \cite{Ciricosta2016} and theoretical \cite{Zeng2020} results are available, and to extend the study to other processes which can be affected by plasma density effects, such as radiative recombination \cite{Ribiere2022}.
\appendix
\section{Integrals}\label{Integrals}

The integrals in Eqs. (\ref{eqn:Il}) and (\ref{eqn:K}) are respectively
\begin{eqnarray*}
    I_1&=&\frac{1}{\sqrt{2\pi\,v}}\frac{1}{E}\int_0^{\infty}E_i\exp\left[-(E_i-\mu_1)^2/(2v)\right]dE_i\nonumber\\
    &=&\frac{1}{\sqrt{2\pi\,v}}\frac{1}{E}\left\{ve^{-\frac{\mu_1^2}{2v}}+\mu_1\sqrt{v}\sqrt{\frac{\pi}{2}}\left[1+\mathrm{erf}\left(\frac{\mu_1}{\sqrt{2v}}\right)\right]\right\},\nonumber\\
    I_2&=&\frac{1}{\sqrt{2\pi\,v}}\frac{1}{E^2}\int_0^{\infty}E_i^2\exp\left[-(E_i-\mu_1)^2/(2v)\right]dE_i\nonumber\\
    &=&\frac{1}{\sqrt{2\pi\,v}}\frac{1}{E^2}\left\{\mu_1ve^{-\frac{\mu_1^2}{2v}}+\sqrt{v}(\mu_1^2+v)\sqrt{\frac{\pi}{2}}\left[1+\mathrm{erf}\left(\frac{\mu_1}{\sqrt{2v}}\right)\right]\right\},\nonumber\\
    I_3&=&\frac{1}{\sqrt{2\pi\,v}}\frac{1}{E^3}\int_0^{\infty}E_i^3\exp\left[-(E_i-\mu_1)^2/(2v)\right]dE_i\nonumber\\
    &=&\frac{1}{\sqrt{2\pi\,v}}\frac{1}{E^3}\left\{v(2v+\mu_1^2)\exp\left(-\mu_1^2/(2v)\right)+\mu_1\sqrt{\frac{\pi v}{2}}(\mu_1^2+3v)\left[1+\mathrm{erf}\left(\frac{\mu_1}{\sqrt{2v}}\right)\right]\right\}
\end{eqnarray*}
and
\begin{eqnarray*}
    K&=&\frac{1}{\sqrt{2\pi\,v}}\frac{1}{E}\int_0^{\infty}E_i\ln E_i\exp\left[-(E_i-\mu_1)^2/(2v)\right]dE_i\nonumber\\
    &=&\frac{1}{\sqrt{2\pi\,v}}\frac{1}{E}\,e^{-\frac{\mu_1^2}{2v}}\left\{-\gamma+\ln(2v)-\mu_1e^{\frac{\mu_1^2}{2v}}\sqrt{\frac{\pi}{2v}}\left(\vphantom{\frac{\mu_1^2}{2v}}-2+\gamma+2\ln 2-\ln(2v)\right.\right.\nonumber\\
    &+&\left.\mathrm{erf}\left(\frac{\mu_1}{\sqrt{2v}}\right)(\gamma-\ln(2v))+~_1F_1^{(1,0,0)}\left[0,\frac{3}{2};-\frac{\mu_1^2}{2v}\right]\right)\left.+~_1F_1^{(1,0,0)}\left[1,\frac{1}{2};\frac{\mu_1^2}{2v}\right]\right\},
\end{eqnarray*}
where $\mathrm{erf}$ is the error function \cite{Abramowitz1964}. $_1F_1^{(1,0,0)}[a,b;z]$ is the derivative with respect to $a$ of the Kummer confluent hypergeometric function $_1F_1[a,b;z]$ \cite{Abramowitz1964,Ancarani2009}:
\begin{equation*}
    _1F_1^{(1,0,0)}\left[a,b;z\right]=\frac{\partial}{\partial a}~_1F_1[a,b;z]\equiv G^{(1)}(a,b;z).
\end{equation*}
We have to evaluate $G^{(1)}$ for $a=0$ and $a=1$. The first case yields:
\begin{equation*}
    ~_1F_1^{(1,0,0)}\left[0,\frac{3}{2};-\frac{\mu_1^2}{2v}\right]= G^{(1)}\left(0,\frac{3}{2};-\frac{\mu_1^2}{2v}\right)=\frac{z}{\left(\frac{3}{2}\right)_1}~_2F_2\left(\begin{array}{l}
    1,1\\
    2,\frac{5}{2}
    \end{array}\left.\right|;z\right),
\end{equation*}
where $_2F_2\left(\begin{array}{l}
1,1\\
2,\frac{5}{2}
\end{array}\left.\right|;z\right)$
is the generalized hypergeometric function which can be replaced by the expansion \cite{Kang2015,Olver2011}:
\begin{equation*}
    _2F_2\left(\begin{array}{l}
    a_1,a_2\\
    b_1,b_2
    \end{array}\left.\right|;z\right)=\sum_{n=0}^{\infty}\frac{\Pi_{i=1}^{2}(a_i)_n}{\Pi_{j=1}^{2}(b_j)_n}\frac{z^n}{n!},
\end{equation*}
where $(x)_n$ is the Pochhammer symbol, defined in terms of the Gamma function as $(x)_n=\Gamma(x+n)/\Gamma(x)$. As the number of $a_i$ terms is equal to the number of $b_j$ ones the expansion converges for all $z$ values \cite{Kang2015}. We also need to calculate $\displaystyle G^{1)}(1,1/2;z)$, where $z=\mu_1^2/(2v)$. After a straightforward calculation (see Ref. \cite{Ancarani2009}), one obtains:
\begin{eqnarray*}
    G^{(1)}[1,1/2;z]&=&\sum_{m_1=1}^{\infty}\frac{(1)_{m_1}}{(1/2)_{m_1}}\frac{z^{m_1}}{m_1!}\sum_{p=0}^{m_1-1}\frac{1}{p+1}\nonumber\\
    &=&\sum_{m_1=1}^{\infty}\frac{m_1!\,2^{2m_1}}{(2m_1)!}z^{m_1}\sum_{p=0}^{m_1-1}\frac{1}{p+1}.
\end{eqnarray*}

\section{Calculations using a representation of the Gaussian}\label{Cubic_splines}

The Gaussian can be sampled at the points $u = -m, -m+ 1, \cdots, 0, \cdots , m-1, m$ (in practice, we take $m=6$) and interpolated using cubic splines \cite{deBoor1978} on each interval $[k, k + 1]$ by the formula
\begin{equation}
    {\rm G}(u)=\frac{1}{\sqrt{2\pi\,v}}\left[a_k+b_k\,u+c_k\,u^2+d_k\,u^3\right].\label{Gauss-distrib}
\end{equation}
The coefficients $a_k$, $b_k$, $c_k$ and $d_k$ in the interval $[k, k + 1]$ are determined by the continuity of the function and its derivative at the points $u = k$ and $u = k + 1$. One then has
\begin{equation}
    {\rm G}(u)=\frac{e^{-(k+1)^2/2}}{\sqrt{2\pi\,v}}\left[\tilde{a}_k+\tilde{b}_k\,u+\tilde{c}_k\,u^2+\tilde{d}_k\,u^3\right],\label{Gauss-distrib2}
\end{equation}
where the coefficients $\tilde{a}_k$, $\tilde{b}_k$, $\tilde{c}_k$ and $\tilde{d}_k$ are given in Table \ref{ak-Gaussian} \cite{Pain2015}. The Gaussian is assumed to be zero for $|u| > m$.

\begin{table}[ht]
\caption{\label{ak-Gaussian}Expressions of the coefficients $\tilde{a}_k$, $\tilde{b}_k$, $\tilde{c}_k$ and $\tilde{d}_k$ as functions of $k$.}
\footnotesize
\centering
\begin{tabular}{|l|l|}
\hline
Coefficient & Expression\\
\hline
$\tilde{a}_k$ & $\left[k^2(k+2)^2+e^{k+1/2}\left(k^2-1\right)^2\right]$ \\
\hline
$\tilde{b}_k$ & $-k(k+1)\left[8+3k+e^{k+1/2}\left(3k-5\right)\right]$ \\
\hline
$\tilde{c}_k$ & $\left[4+10k+3k^2+e^{k+1/2}\left(3k^2-4k-3\right)\right]$ \\
\hline
$\tilde{d}_k$ & $-\left[3+e^{k+1/2}(k-2)+k\right]$ \\
\hline
\end{tabular}\label{ak}
\end{table}

Using Eqs. (\ref{New-definition})-(\ref{XS-Kim}) and (\ref{Gauss-distrib})-(\ref{Gauss-distrib2}) with Table \ref{ak-Gaussian}, we can write the cross sections as
\begin{eqnarray*}
    \sigma(E)&=&\frac{1}{\sqrt{2\pi\,v}}\int_0^{\infty}\sigma(E|E_i)\exp\left[-(E_i-\mu_1)^2/(2v)\right]dE_i\nonumber\\
    &=&\frac{1}{\sqrt{2\pi\,v}}\sum_{k=0}^{\infty}\int_{\epsilon_k}^{\epsilon_{k+1}}\left[A\frac{\ln(E/E_i)}{E/E_i}+\sum_{l=1}^3\frac{B_l}{(E/E_i)^l}\right]\nonumber\\
    & &\times\left[a_k+b_k\left(\frac{E_i-\mu_1}{\sqrt{v}}\right)\right.\left.+c_k\left(\frac{E_i-\mu_1}{\sqrt{v}}\right)^2+d_k\left(\frac{E_i-\mu_1}{\sqrt{v}}\right)^3\right]dE_i.
\end{eqnarray*}
If we first consider the terms which do not involve the logarithm we are left with integrals of the type
\begin{equation*}
    F_{l,n,k}=\frac{1}{E^l}\frac{1}{v^{n/2}}\int_{\epsilon_k}^{\epsilon_{k+1}} E_i^l\left(E_i-\mu_1\right)^n\,dE_i,
\end{equation*}
where $l$ and $n$ are integers: $1\leq l\leq 3$ and $0\leq n\leq 3$. 
We can then write their contribution to the cross section as
\begin{equation*}
    \sigma_1(E)=\frac{1}{\sqrt{2\pi\,v}}\sum_{k=0}^{\infty}\left[a_k \sum_{l=1}^3B_l\,F_{l,0,k}+b_k \sum_{l=1}^3B_l\,F_{l,1,k}+c_k \sum_{l=1}^3B_l\,F_{l,2,k}+d_k \sum_{l=1}^3B_l\,F_{l,3,k}\right].
\end{equation*}
The remaining contribution involves the following integrals:
\begin{equation*}
    J_{n,k}=\frac{\ln\,E}{E}\frac{1}{v^{n/2}}\int_{\epsilon_k}^{\epsilon_{k+1}}E_i(E_i-\mu_1)^n\,dE_i
\end{equation*}
and
\begin{equation*}
    K_{n,k}=-\frac{1}{E}\frac{1}{v^{n/2}}\int_{\epsilon_k}^{\epsilon_{k+1}}E_i\ln(E_i)(E_i-\mu_1)^n\, dE_i.
\end{equation*}
Setting $\displaystyle L_{n,k}=J_{n,k}+K_{n,k}$, the corresponding cross section $\sigma_2$ reads
\begin{equation*}
    \sigma_2=\frac{1}{\sqrt{2\pi\,v}}\sum_{k=0}^{\infty}\left[a_kL_{0,k}+b_kL_{1,k}+c_kL_{2,k}+d_kL_{3,k}\right].
\end{equation*}
We have
\begin{eqnarray*}
    F_{l,n,k}&=&\frac{1}{E^l}\frac{1}{v^{n/2}}\int_{\epsilon_k}^{\epsilon_{k+1}} E_i^l(E_i-\mu_1)^n\,dE_i\\
    &=&\frac{1}{E^l}\,\frac{1}{v^{n/2}}\sum_{p=0}^n\binom{n}{p}\,(-\mu_1)^{n-p}\,\int_{\epsilon_k}^{\epsilon_{k+1}}E_i^l\,E_i^p\,dE_i\\
    &=&\frac{1}{E^l}\,\frac{1}{v^{n/2}}\sum_{p=0}^n\binom{n}{p}(-\mu_1)^{n-p}\left[\frac{E_i^{l+p+1}}{l+p+1}\right]_{\epsilon_k}^{\epsilon_{k+1}},
\end{eqnarray*}
which can be written as
\begin{equation*}
    F_{l,n,k}=\frac{1}{E^l}\,\frac{1}{v^{n/2}}\sum_{p=0}^n\binom{n}{p}(-\mu_1)^{n-p}\frac{1}{l+p+1}\left[\epsilon_{k+1}^{l+p+1}-\epsilon_k^{l+p+1}\right].
\end{equation*}
Similar calculations yield
\begin{equation*}
    J_{n,k}=\frac{\ln\,E}{E}\,\frac{1}{v^{n/2}}\sum_{p=0}^n\binom{n}{p}(-\mu_1)^{n-p}\frac{1}{p+2}\left[\epsilon_{k+1}^{p+2}-\epsilon_k^{p+2}\right]
\end{equation*}
and
\begin{equation*}
    K_{n,k}=-\frac{1}{E}\,\frac{1}{v^{n/2}}\sum_{p=0}^n\binom{n}{p} (-\mu_1)^{n-p}\int_{\epsilon_k}^{\epsilon_{k+1}}\ln(E_i)\,E_i^{p+1}\,\,dE_i.
\end{equation*}
Using
\begin{equation*}
    \int_{\epsilon_k}^{\epsilon_{k+1}}y^p\,\ln\,y\,dy=\frac{[1-(p+1)\ln(\epsilon_k)]\epsilon_k^{p+1}-[1-(p+1)\ln(\epsilon_{k+1})]\epsilon_{k+1}^{p+1}}{(p+1)^2},
\end{equation*}
we obtain
\begin{equation*}
    K_{n,k}=-\frac{1}{E}\,\frac{1}{v^{n/2}}\sum_{p=0}^n\binom{n}{p} (-\mu_1)^{n-p}\left[\frac{[1-(p+2)\ln(\epsilon_k)]\epsilon_k^{p+2}-[1-(p+2)\ln(\epsilon_{k+1})]\epsilon_{k+1}^{p+2}}{(p+2)^2}\right].
\end{equation*}

%\clearpage

\end{document}